\documentclass[onecolumn]{aa}  
\usepackage{mathtools}
\usepackage{graphicx}
\usepackage{natbib}
\usepackage{color} 
\usepackage{txfonts}
\bibpunct{(}{)}{;}{a}{}{,} 
\newcommand{\nc}{\newcommand}

\newcommand{\CII}{[C\,{\sc ii}]}

\newcommand{\CI}{[C\,{\sc i}]} 

\newcommand{\NII}{[N\,{\sc ii}]}

\newcommand{\OI}{[O\,{\sc i}]}
\newcommand{\cplus}{C$^+$}
\newcommand{\HII}{H\,{\sc ii}}
\newcommand{\HI}{H\,{\sc i}}

\newcommand{\Halpha}{H$\alpha$}
\nc\micron{\mbox{$\mu$m}}
\nc{\cmcub}{\mbox{cm$^{-3}$}}
\nc{\cmsq}{\mbox{cm$^{-2}$}}
\nc{\Kkms}{\mbox{K~km~s$^{-1}$}}
\nc{\kms}{\mbox{km~s$^{-1}$}}
\nc{\mthirty}{\mbox{M\,33}}
\nc{\Tmb}{\mbox{$T_{\rm mb}$}}
\nc{\Tkin}{\mbox{$T_{\rm kin}$}}
\nc{\vlsr}{\mbox{v$_{\rm LSR}$}}
\newcommand\arcdeg{\mbox{$^\circ$}}%
\nc{\twCO}{$^{12}$CO}
\nc{\thCO}{$^{13}$CO}
\nc{\msun}{\ensuremath{\mathrm{M}_\odot}}
\nc{\rsun}{\ensuremath{\mathrm{R}_\odot}}
\nc{\lsun}{\ensuremath{\mathrm{L}_\odot}}
\begin{document}
\title{Velocity resolved \CII\ spectroscopy of  the center and the 
BCLMP\,302 region of M\,33
(HerM\,33es)
  \thanks{Herschel is an ESA
    space observatory with science instruments provided by
    European-led Principal Investigator consortia and with important
    participation from NASA.\fnmsep All spectra are uploaded as ascii
    files on CDS.}}

\titlerunning{HIFI spectroscopy of M\,33 center \& BCLMP302}

   \author{B. Mookerjea \inst{\ref{tifr}}, 
   F.  Israel \inst{\ref{leiden}}, 
   C.  Kramer \inst{\ref{iramsp}}, 
   T.  Nikola\,\inst{\ref{cornell}}, 
   J. Braine \inst{\ref{bordeaux}}, 
   V. Ossenkopf \inst{\ref{kosma}},
   M. R\"ollig \inst{\ref{kosma}},
   C. Henkel \inst{\ref{mpifr},\ref{saudi}},
   P. van der Werf \inst{\ref{leiden}},
   F. van der Tak \inst{\ref{kapteyn},\ref{sron}},
   M. C. Wiedner \inst{\ref{obspm}}}
\institute{Tata Institute of Fundamental Research, Homi Bhabha Road,
Mumbai 400005, India \email{bhaswati@tifr.res.in}\label{tifr}
\and
Leiden Observatory, Leiden University, PO Box 9513, NL 2300
RA Leiden, The Netherlands \label{leiden}
\and
Instituto Radioastronom\'{i}a Milim\'{e}trica, Av. Divina Pastora 7,
Nucleo Central, E-18012 Granada, Spain \label{iramsp}
\and
Department of Astronomy, Cornell University, Ithaca, NY 14853 \label{cornell}
\and
Laboratoire d'Astrophysique de Bordeaux, Universit\'{e} Bordeaux
1, Observatoire de Bordeaux, OASU, UMR 5804, CNRS/INSU, B.P.
89, Floirac F-33270 \label{bordeaux}
\and
KOSMA, I. Physikalisches Institut, Universit\"at zu K\"oln,
Z\"ulpicher Strasse 77, D-50937 K\"oln, Germany \label{kosma}
\and
Max Planck Institut f\"ur Radioastronomie, Auf dem H\"ugel 69, D-
53121 Bonn, Germany \label{mpifr}
\and
Astron. Dept., King Abdulaziz University, P.O. Box 80203,
  Jeddah 21589, Saudi Arabia \label{saudi}
\and
Kapteyn Astronomical Institute, University of Groningen, PO box 800,
9700 AV Groningen, The Netherlands \label{kapteyn}
\and
SRON Netherlands Institute for Space Research, Landleven 12, 9747 AD
Groningen, The Netherlands \label{sron}
\and
LERMA, Observatoire de Paris, PSL Research University, CNRS, Sorbonne
Universités, UPMC Univ. Paris 06, F-75014, Paris, France
\label{obspm}
}

\date{Received \ldots; accepted \ldots}

\authorrunning{Mookerjea et al.}

\abstract
{The forbidden fine structure transition of \cplus\ at 158\,\micron\ is
one of the major cooling lines of the interstellar medium (ISM). }
{We aim to understand the contribution of the ionized, atomic and
molecular phases of the ISM to the \CII\ emission from clouds near the
dynamical center and the BCLMP302 \HII\ region in the north  of the
nearby galaxy M\,33  at a spatial resolution of 50\,pc.}
{We combine high resolution \CII\ spectra taken with the HIFI
spectrometer onboard the Herschel satellite with \CII\ Herschel-PACS
maps and ground-based observations of CO(2-1) and \HI.  All data 
are at a common spatial resolution of 50\,pc.  Correlation coefficients between the integrated
intensities of \CII, CO(2--1) and \HI\ are estimated from the
velocity-integrated PACS data and from the HIFI data.
 We decomposed the \CII\ spectra in terms of contribution
from molecular and atomic gas detected in CO(2--1) and \HI,
respectively. At a few positions, we estimated the contribution of ionized
gas to \CII\ from the emission measure observed at radio wavelengths.}
{In both regions, the center and BCLMP302, the correlation seen in the
\CII, CO(2--1) and \HI\ intensities from structures of all sizes is
significantly higher than the highest correlation in intensity obtained
when comparing only structures of the same size.  The correlations
between the intensities of tracers corresponding to the same velocity
range as \CII, differ from the correlation derived from PACS data.
Typically, the \CII\ lines have widths intermediate between the narrower
CO(2--1) and broader \HI\ line profiles. A comparison of the spectra
shows that
the relative contribution of molecular and atomic gas traced by CO(2--1)
and \HI\ varies substantially between positions and depends mostly on
the local physical conditions and geometry. At the positions of the
\HII\ regions, the ionized gas contributes between 10--25\% of the
observed \CII\ intensity.  We estimate that 11--60\% and 5--34\% of the
\CII\ intensities in the center and in BCLMP302, respectively, arise at
velocities showing no CO(2--1) or \HI\ emission and could arise in
CO-dark molecular gas.  The deduced strong variation in the \CII\
emission  not associated with CO and \HI\ cannot be explained in terms
of differences in $A_{\rm V}$, far-ultraviolet radiation field, and
metallicity between the two studied regions.  }
{The relative amounts of diffuse (CO-dark) and dense molecular gas
possibly vary on spatial scales smaller than 50\,pc. In both regions, a
larger fraction of the molecular gas is traced by \CII\ than by the
canonical tracer CO.  Correlations between observed intensities of \CII,
CO, and \HI\ crucially depend on the spatial and spectral resolution of
the data and need to be used carefully, in particular, for extragalactic
studies.  These results emphasize the need for velocity-resolved
observations to discern the contribution of different components of the
ISM to \CII\ emission.}


\maketitle

%

\section{Introduction}

Under a wide range of astrophysical conditions, including ionized,
atomic, and molecular phases of the interstellar medium (ISM), a
significant fraction of gaseous carbon is in the form of \cplus. The
forbidden fine structure transition
$^3$P$_{3/2}$$\rightarrow$$^3$P$_{1/2}$ of \cplus\ (with
$\Delta$E=91\,K), written as \CII, is easily excited and is typically
optically thin and not affected by extinction in most astrophysical
environments. Thus, \CII\ is not only an excellent coolant of the neutral
gas exposed to the ionizing flux of nearby early-type stars, but also an
excellent probe of stellar radiation fields and their effects on the
physical conditions of the ambient neutral gas.

Early studies of \CII\ emission from nearby galaxies revealed that the
line contributes 0.1\% to 1\% of the total far-infrared (FIR) continuum
and is strongly correlated with the strength of the CO(1--0) rotational
line \citep{crawford1985,stacey1991,madden1993,malhotra2001}. These
studies
suggested that \CII\ line emission on galactic scales mostly originates
from the warm, dense photodissociated surfaces of clumps in molecular
clouds, known as photodissociation regions (PDRs).  The PDRs are created
by far-UV (FUV; 6\,eV$<$h$\nu<$13.6\,eV) photons from nearby OB stars or
the general interstellar radiation field (ISRF).  Moderate density
($n_{\rm H}< 10^4$\,\cmcub) and temperature (30 -- 100\,K) clouds in the
PDRs are primarily cooled by the 158\,\micron\ \CII\  line, with \OI\ at
63\,\micron\ being responsible for the cooling at high FUV fields and
large gas densities.  However, high spatial resolution ($<$300\,pc)
observations of nearby galaxies that resolve individual giant molecular
clouds (GMCs) show the correlation between \CII\ and CO(1--0) to be
much weaker \citep{rodriguez2006,mookerjea2011,kramer2013}.  The question of
disentangling the contributions of the components  (PDR, cold neutral
medium (CNM), and ionized gas) to the \CII\ emission from galaxies can
only be
appropriately addressed via observations of several emission
lines at high spatial and spectral resolution. 

In the absence of multiple transitions of \cplus, comparison of spectral
profiles with those of CO, \HI\ and other tracers at resolutions
comparable to ground-based single dish telescopes are crucial and the
velocity-resolved high spatial resolution capabilities of the recent
far-infrared missions HIFI/Herschel and GREAT/Stratospheric Observatory
for Far Infrared Astronomy (SOFIA)  have made these
possible. Located at a distance of 840\,kpc \citep{freedman1991}, the
Triangulum galaxy M\,33 is a nearby, gas-rich disk galaxy with
half the solar metallicity, which allows for this type of  a coherent survey at
high spatial and spectral resolution.  As part of the Herschel Key
Program HerM\,33es \citep{kramer2010}, selected regions along the major
axis of M\,33 were mapped with PACS in \CII, \OI\ (63\,\micron) and
\NII\ (122\,\micron) \citep[][;Nikola et al.  (in prep)]{mookerjea2011}.
These PACS regions are further probed along selected cuts with
velocity-resolved spectra of \CII\ at 158\,\micron\ using HIFI
\citep{braine2012}.  \citet{higdon2003} performed a detailed study
of the nucleus of M\,33 and six selected \HII\ regions using seven FIR
fine-structure lines observed with ISO/LWS.  The present {\em Herschel}
observations also follow up on the ISO/LWS observations of \CII\
integrated intensities at 280\,pc resolution along the major axis of
M\,33 \citep{kramer2013}, encompassing the regions observed in \CII\
with {\em Herschel}.  In this paper, we present higher resolution \CII\
lines observed with HIFI in the central region of M\,33 and the \HII\
region BCLMP302 along two 140\arcsec\ long strips approximately oriented
along the north-south and east-west direction in each of the two
regions.  The BCLMP302 region is located at a galactocentric distance of
2\,kpc in the northern spiral arm.  The HIFI spectra sample multiple
emission features detected in CO and dust continuum emission, with sizes
typically corresponding to GMCs.  The HIFI \CII\ data are compared with
CO(2--1) and \HI\ spectra tracing the molecular and atomic components of
the neutral interstellar medium \citep{gratier2010}.  The primary aim of
the paper is to utilize the velocity information in the newly observed
HIFI \CII\ data to identify the phase of the ISM that mainly contributes
to the \CII\ emission and also to derive its physical properties.  The
work presented here extends the \CII\ PACS and HIFI (single-position)
observations of BCLMP\,302 \citep{mookerjea2011}.  It also compares the
properties of the \CII\ emitting regions in the central region of M\,33
and BCLMP302 with the \HII\ region BCLMP691, located approximately at a
galactocentric distance of 3.5\,kpc  \citep{braine2012}.


\section{Observations}

\subsection{Herschel: \CII\ with HIFI }

We have observed the $^3$P$_{3/2}$--$^3$P$_{1/2}$ transition of \cplus\ at
1900536.9\,MHz with the Heterodyne Instrument for Far Infrared
astronomy (HIFI) onboard Herschel \citep{pilbratt2010,degraauw2010}.
The observations consisted of two load-chopped, on-the-fly (OTF) scans
each of $\sim 140$\arcsec\ length with position angle (PA) of
22.5\arcdeg\ and 112.5\arcdeg\ for an area around the  center of M\,33
and with PA of 45\arcdeg\ and 135\arcdeg\ for a region around the \HII\
region BCLMP302 in M\,33.  The measurements presented here correspond to
the {\em Herschel} observation identification (Obsids) 1342213340
(OD623), 1342213702 (OD630), 1342213741 (OD632) and 1342214315 (OD642).
The observations were performed on 26 January, 2, 4, and 14 February,
2011. The offsets for the OTF positions in the central region are given
relative to the position $\alpha_{2000}$ = 01$^{\rm h}$33$^{\rm
m}$48\fs20 $\delta_{2000}$ = 30\arcdeg39\arcmin21\farcs4  with the
east-west (EW) cut passing through the dynamical center of M\,33. For
the BCLMP302 region, the corresponding offsets are given relative to
$\alpha_{2000}$ = 01$^{\rm h}$34$^{\rm m}$06\fs30 $\delta_{2000}$ =
30\arcdeg47\arcmin25\farcs3. At the frequency of \CII, the Herschel/HIFI
beam size is 11\farcs1, the forward efficiency $F_{\rm eff}$ is 96\%,
and the main beam efficiency $B_{\rm eff}$  is 59\% \citep{mueller2014}. 

The HIFI Wide Band Spectrometer (WBS) was used with spectral resolution
of 1.1\, MHz (0.17\,\kms) over an instantaneous bandwidth of 2.4\,GHz at
1.9\,THz. The \CII\ spectra were smoothed to a resolution of 1.2\,\kms\
to improve the signal-to-noise ratios per channel. 

Data was reduced using the standard HIFI pipeline up to level 2
including HifiFitFringe, with the ESA supported-package HIPE 10.0
\citep{ott2010}. About 30\% of the observed \CII\ spectra were found to
suffer from the standing wave pattern seen in receivers with hot
electron bolometers (HEB), which were corrected to a large extent 
via a tool employing the current matching technique provided by the
Herschel helpdesk. The data were then exported as FITS files into
CLASS/GILDAS format\footnote{http://www.iram.fr/IRAMFR/GILDAS} for
subsequent reduction. The spectra on the two fully-sampled cuts for
each region were averaged over 15\arcsec\ and gridded into a 10\arcsec\
grid (Fig.\,\ref{fig_specplot}).  We obtained the main beam
brightness temperature, $T_{\rm mb}$, by multiplying the antenna
temperature, $T_{\rm A}^\ast$, by
$F_{\rm eff}$/$B_{\rm eff}$.

The gridded spectra from the central region of M\,33, which we selected
for further analysis, have typical integration times between 20 to 40
minutes and show an rms of 48 to 62\,mK for a velocity resolution of
1.2\,\kms.  Similarly, for BCLMP302 at a velocity resolution of
1.2\,\kms\ the final rms noise varies between 42 to 54\,mK with most
positions having an integration time of 54 mins.  The rms levels
mentioned above are in the $T_{\rm mb}$ scale.

\subsection{Complementary data}

The PACS/Herschel spectroscopy observations of \CII, [OI] 63\,\micron, and
\NII\ 122\,\micron\ contain 22 (including BCLMP302 and the center)
independent observations of smaller regions, each covered by a 3x3
raster with the PACS array, and combined they result in a strip map
along the major axis of M33. In total, these observations cover a region
of about 45\arcmin\ $\times$ 2\farcm5 at a resolution of 12\arcsec\
with some gaps in the northern part. The complete PACS spectroscopic
dataset will be presented in Nikola et al. (in prep).  The \HI\
VLA and CO(2--1) HERA/30m spectra at the HIFI \CII\ positions were
extracted from the maps presented by \citet{gratier2010}. The original
angular resolutions of the \HI\ and CO(2--1) data were 5\arcsec\ and
12\arcsec, respectively. The spectral resolutions of the \HI\ and
CO(2--1) data were 2.6\,\kms\ and 1.3\,\kms, respectively.  We also use
the {\it Spitzer} MIPS 24\,\micron\ data presented by
\citet{tabatabaei2007}, and the {\sc HerM\,33es} PACS and SPIRE maps at
100, 160, 250, 350 and 500\,\micron\ \citep{verley2010, boquien2010,
kramer2010,boquien2011,xilouris2012}.  All complementary Herschel data
up to 160\,\micron\ were smoothed to a resolution of 12\arcsec\ for
direct comparison with \CII.

\section{Global morphology of \CII\ emission in M\,33}
\begin{figure}[h]
\centering
\includegraphics[width=0.48\textwidth,angle=0]{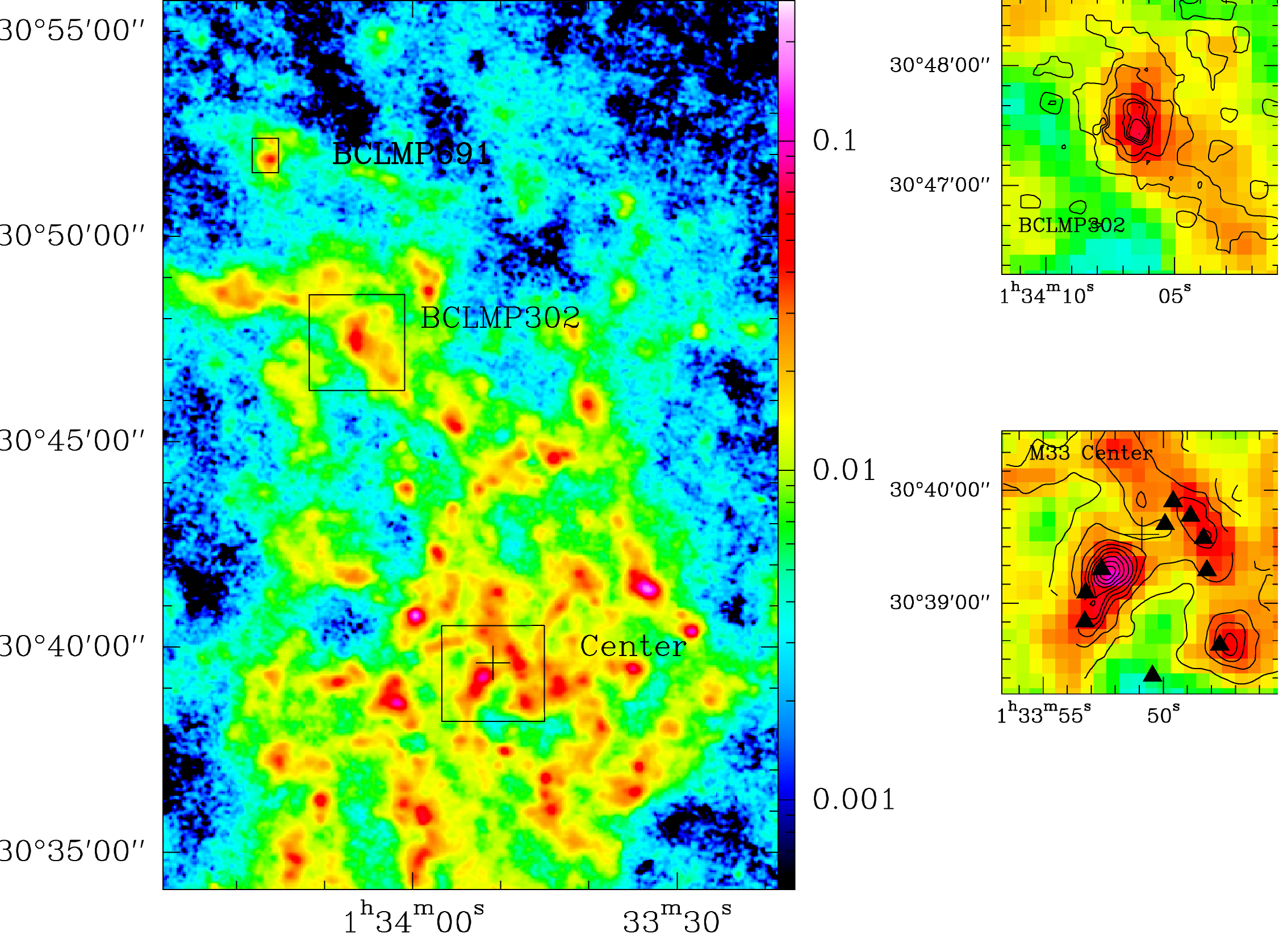}
\caption{PACS 160\,\micron\ continuum image of M\,33 \citep{kramer2010}.
The boxes corresponding to the regions around the \HII\
regions BCLMP302, BCLMP691 and the center of M\,33 are indicated. The cross shows the
position of the dynamical center of M\,33. Zoomed in images of the
two regions studied, the center and BCLMP302, are
shown  to the right, with contours
indicating the integrated \CII\ intensities observed with PACS
(\citet{mookerjea2011}; Nikola et al. in prep). The known 
\HII\ regions in the central region \citep{gordon1999} are shown as filled 
triangles.}
\label{fig_overview}
\end{figure}

Figure~\ref{fig_overview} shows the 160~\micron\ PACS continuum map of
M\,33 with squares indicating the central, BCLMP302, and BCLMP691
regions.  Contours of \CII\ intensity maps observed with PACS for the
two regions presented here (central and BCLMP302) are overlayed on the
160\,\micron\ continuum map and are shown in the insets of
Fig.\,\ref{fig_overview}.  For both regions, the \CII\ integrated
intensity contours follow the dust continuum distribution rather
closely. Toward the center and in BCLMP302 region the \CII\ intensity
maps are also morphologically similar to the tracers of star formation
such as the H$\alpha$ emission. The ISO/LWS \CII\ study along the major
axis also shows a relatively constant correlation with the H$\alpha$ and
the far-infrared continuum; the $I_{\rm CII}$/$I_{\rm FIR}$ ratio stays
constant at 0.8\% in the inner 4.5\,kpc of M\,33, rising in the
outskirts to values of 3\% at 6\,kpc radial distance \citep{kramer2013}.

\begin{figure*}[h]
\centering
\includegraphics[width=0.50\textwidth,angle=0]{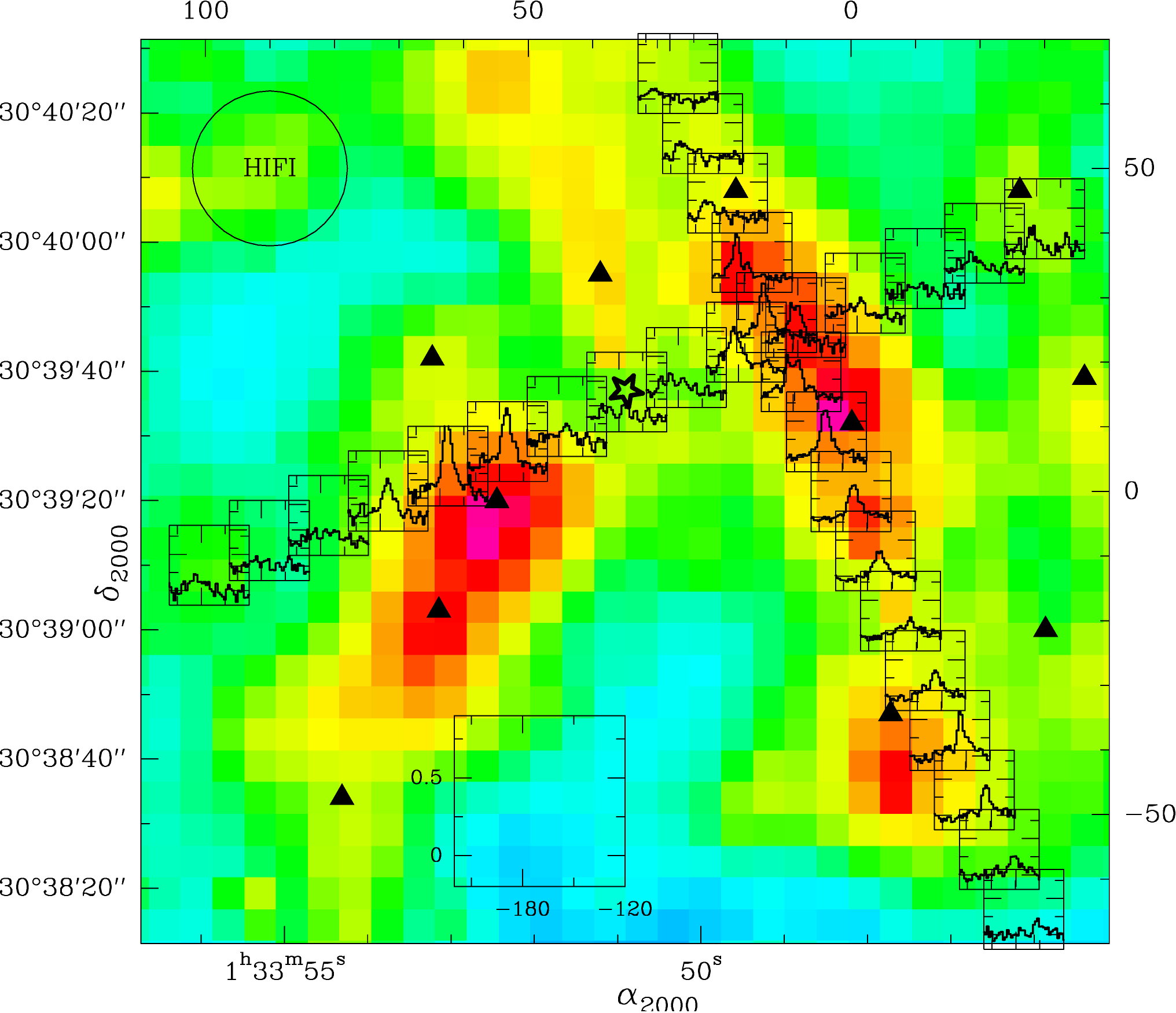}
\includegraphics[width=0.48\textwidth,angle=0]{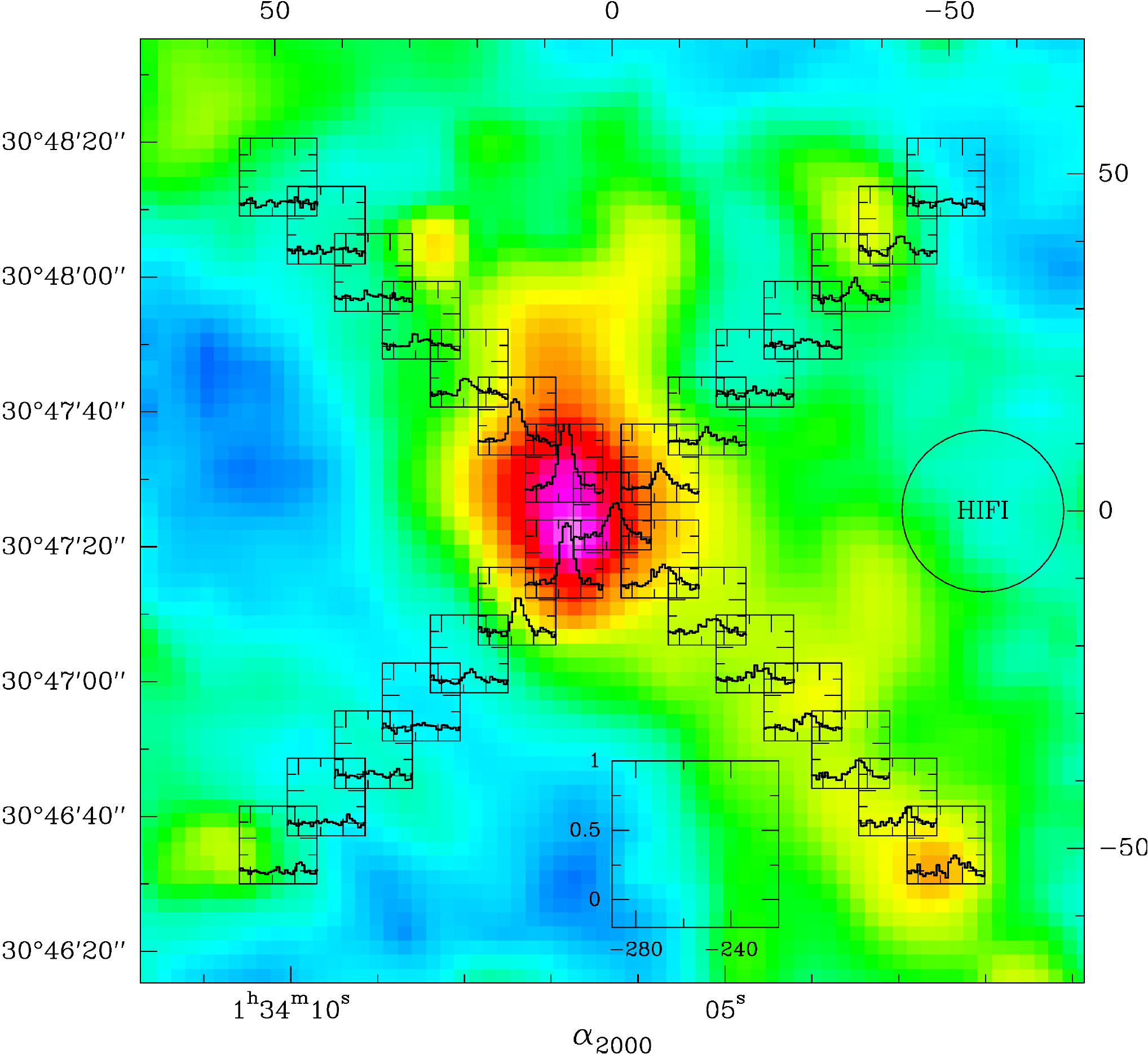}
\caption{MIPS 24~\micron\ image of the regions around the center of
M\,33 (left) and the \HII\ region BCLMP302 (right)  in  M\,33
overplotted with the observed spectra of \CII.   Filled black triangles 
in the {\em
left} box denote positions of the GMCs identified by \citet{gratier2012}
and the star indicated on the eastern part of the E-W cut shows
the position  of the dynamical center of M\,33. The HIFI beam at
1.9\,THz is also shown as inset. All spectra are shown on a 10\arcsec\
grid. Empty boxes show the intensity and velocity scales for the spectra
plotted.
\label{fig_specplot}}
\end{figure*}

Figure\,\ref{fig_specplot} shows the observed \CII\ spectra at 29
positions on a 10\arcsec\ grid along the two cuts observed in each of
the two regions in the center of M\,33 and around BCLMP302, overlayed on
the MIPS 24\,\micron\ image.  Out of the 29 observed positions in each
of the regions, \CII\ emission has been detected at 23 and 19
positions in the central and BCLMP302 regions, respectively.

\subsection{M\,33 central region}

The dynamical center \citep[$\alpha_{2000}$=01$^h$33$^m$50\fs9
$\delta_{2000}$=30\arcdeg39\arcmin35.8\arcsec;][]{jarrett2003} of M\,33
is unspectacular in both \CII\ and dust continuum, with no emission
feature located exactly at that position. However, two intersecting
ridges, consisting of bright infrared peaks, are discernible and the HIFI
cuts are approximately oriented along the two ridges.  Within the
observed region at the center of M\,33, 12 GMCs
\citep[\# 95, 99, 100, 107, 108, 109, 110, 111, 114, 116, 175, and
176;][]{gratier2012} and ten \HII\ regions \citep{gordon1999} are
already identified (Figs.\,\ref{fig_overview},\ref{fig_specplot}). 

The cuts along which the \CII\ HIFI observations are performed, extend
approximately along the north-south (N-S) and the east-west (E-W)
directions (Fig.\,\ref{fig_specplot}).  The N-S cut passes through
several bright continuum sources, while the E-W cut traces the outskirts of
a few of the continuum sources.  Because of the position of the cuts
across the dynamical center of M\,33, the \vlsr\ of peak \CII\ emission
also changes by up to 40\,\kms\ along the two cuts (Sec. 4).  The E-W cut
passes through the dynamical center of M\,33 and we do not detect any
\CII\ emission at the center with HIFI.  The \CII\ intensity measured by
PACS at the center of M\,33 is 5.9\,K\,\kms, which for the typical
velocity spread of 50\,\kms\ of the \CII\ line in this region
corresponds to a peak $T_{\rm mb}$ of 110\,mK.  The rms measured in the
HIFI spectrum at the position of the center at a resolution of
1.2\,\kms\ is 71\,mK. Thus the non detection of \CII\ emission at the
center in the HIFI data is consistent with the PACS observations.
Depending on the position of the HIFI \CII\ beam relative to the
location of molecular clouds, the \CII\ spectra exhibit one or more
velocity components. In addition, the \CII\ intensities decrease with
increasing distance from the dust continuum peaks.

\subsection{BCLMP302}

The mid- and far-infrared dust continuum emission from the BCLMP302
region is dominated by the \HII\ region itself and extends along the
NE-SW direction with a secondary peak approximately 75\arcsec\ to the SW
of the main peak (Fig.\,\ref{fig_specplot}). Previous PACS observations
have shown that the \CII\ intensities in this region correlate well with
the dust continuum and H$\alpha$ emission, thus broadly tracing star-
forming regions \citep{mookerjea2011}. However, the \CII\ and \OI\
63\,\micron\ intensities do not correlate well with either the CO(2-1)
or the \HI\ emission.  One of the two HIFI cuts is aligned with the dust
continuum emission ridge extending in the NE-SW direction and passes
through the secondary peak to the SW.  We detected \CII\ emission primarily
close to the dust continuum peaks,  although we see no intensity
enhancement toward the SW peak. The spectra typically show a
single velocity component except at a few positions immediately to the
SW of the main dust continuum peak, where an additional  narrow velocity
component is also detected.

\section{Integrated \CII\ intensities}
\subsection{Spatial distribution \label{sec_pacs}}

\begin{figure}[h]
\centering
\includegraphics[width=0.42\textwidth,angle=0]{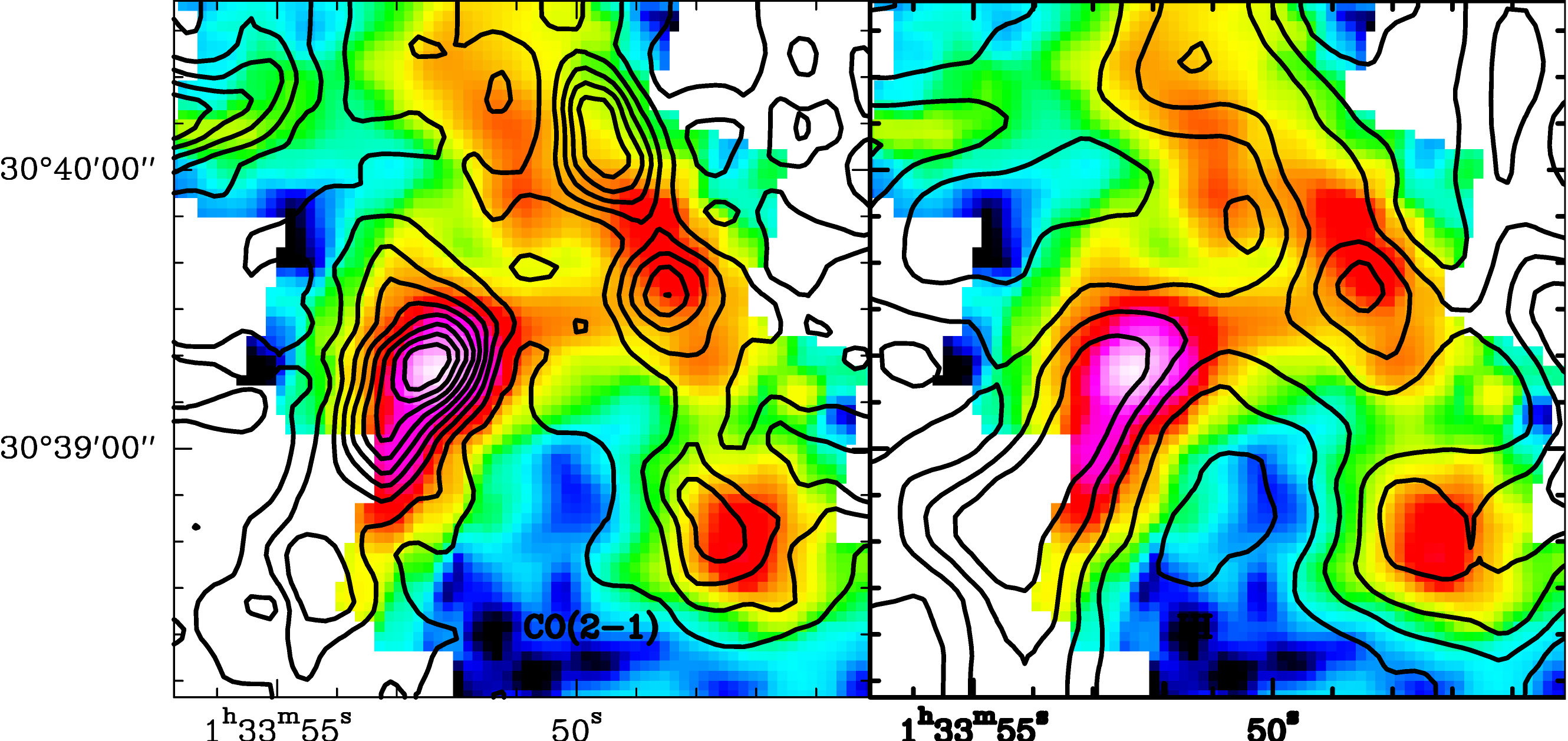}\\
\vspace*{0.5cm}
\includegraphics[width=0.41\textwidth,angle=0]{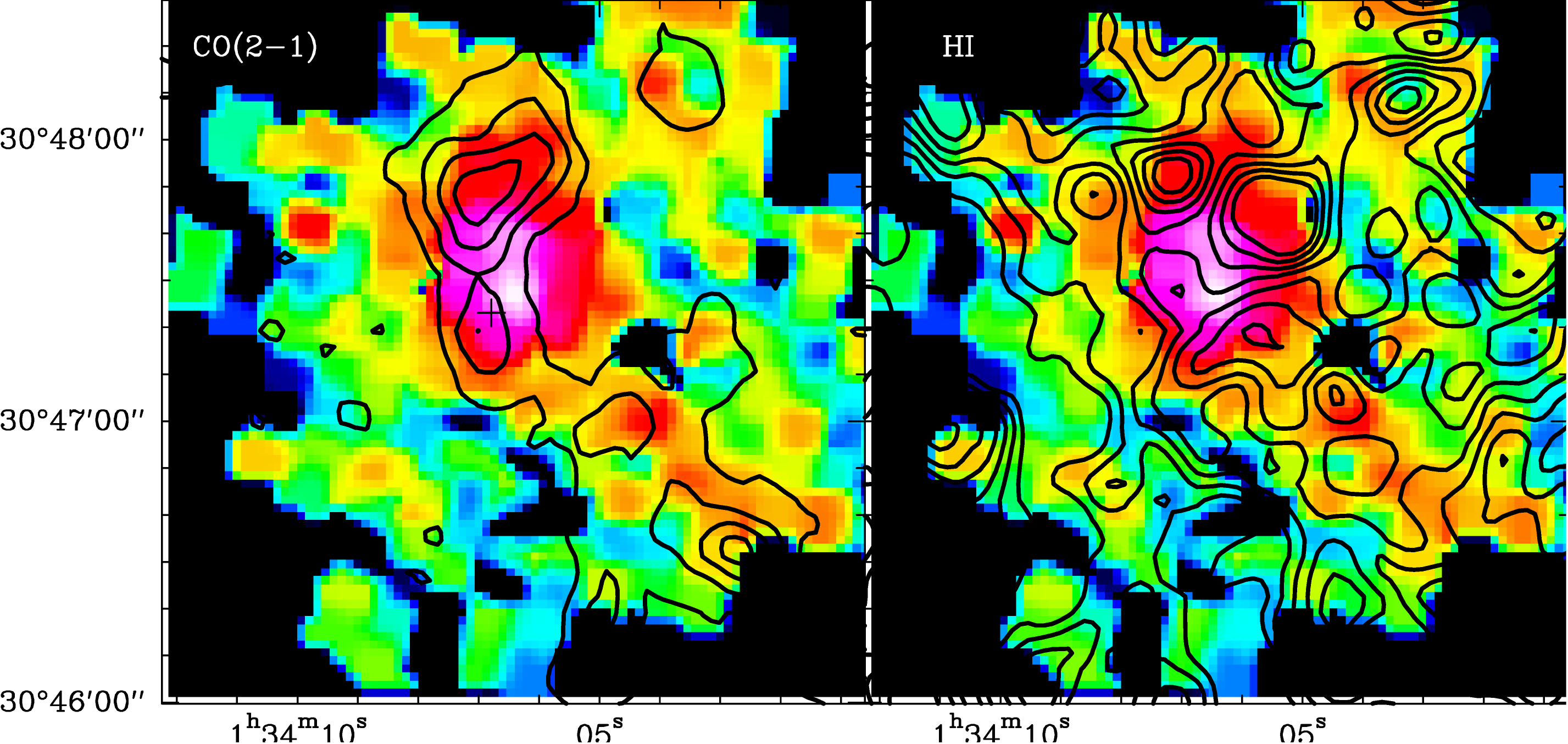}
\caption{Overlay of CO(2--1) ($left$) \& \HI\ ($right$) contours on the
PACS intensity map of \CII\ for the central ($top$) and BCLMP302
($bottom$; \citet{mookerjea2011}) regions. The \CII\ data has no velocity information and CO(2--1)
and \HI\ spectra have been integrated over the entire range of
velocities. The white and black regions in the top and bottom panels 
respectively, indicate regions not mapped with PACS.
\label{fig_pacscomp}}
\end{figure}

Figure\,\ref{fig_pacscomp} shows a comparison of the maps of integrated
intensities of \CII\ (observed with PACS), CO(2--1), and \HI\ for the
central region (top panel) of M\,33 and BCLMP302 (bottom panel).
Correspondence between the emission from the three tracers is typically
considered to be an indicator of the physical proximity or association
of the emitting atomic, molecular, and PDR gas. In the central region, the
main emission features of \CII\ agree in position with CO(2--1) and \HI,
and also with tracers of star formation such as the 160 and 24\,\micron\
dust continuum or H$\alpha$. 

\citet{mookerjea2011} used the PACS \CII\ data to perform an analysis of
the linear correlation of the integrated intensities of CO(2--1) and
\HI\ relative to those of \CII\ for the BCLMP302 region.  These authors
derived linear correlation coefficients of 41\% and 22\%  for \CII--CO
and \CII--\HI, respectively,  using the same maps as in
Fig.\,\ref{fig_pacscomp}. We perform a similar analysis of the \CII\
PACS data for the mapped region around the center of M\,33 and find the
global correlation coefficients to be 71\% and 69\% for the
\CII--CO(2--1) and \CII--\HI\ integrated intensities, respectively.

\begin{figure}[h]
\centering
\includegraphics[width=0.40\textwidth,angle=0]{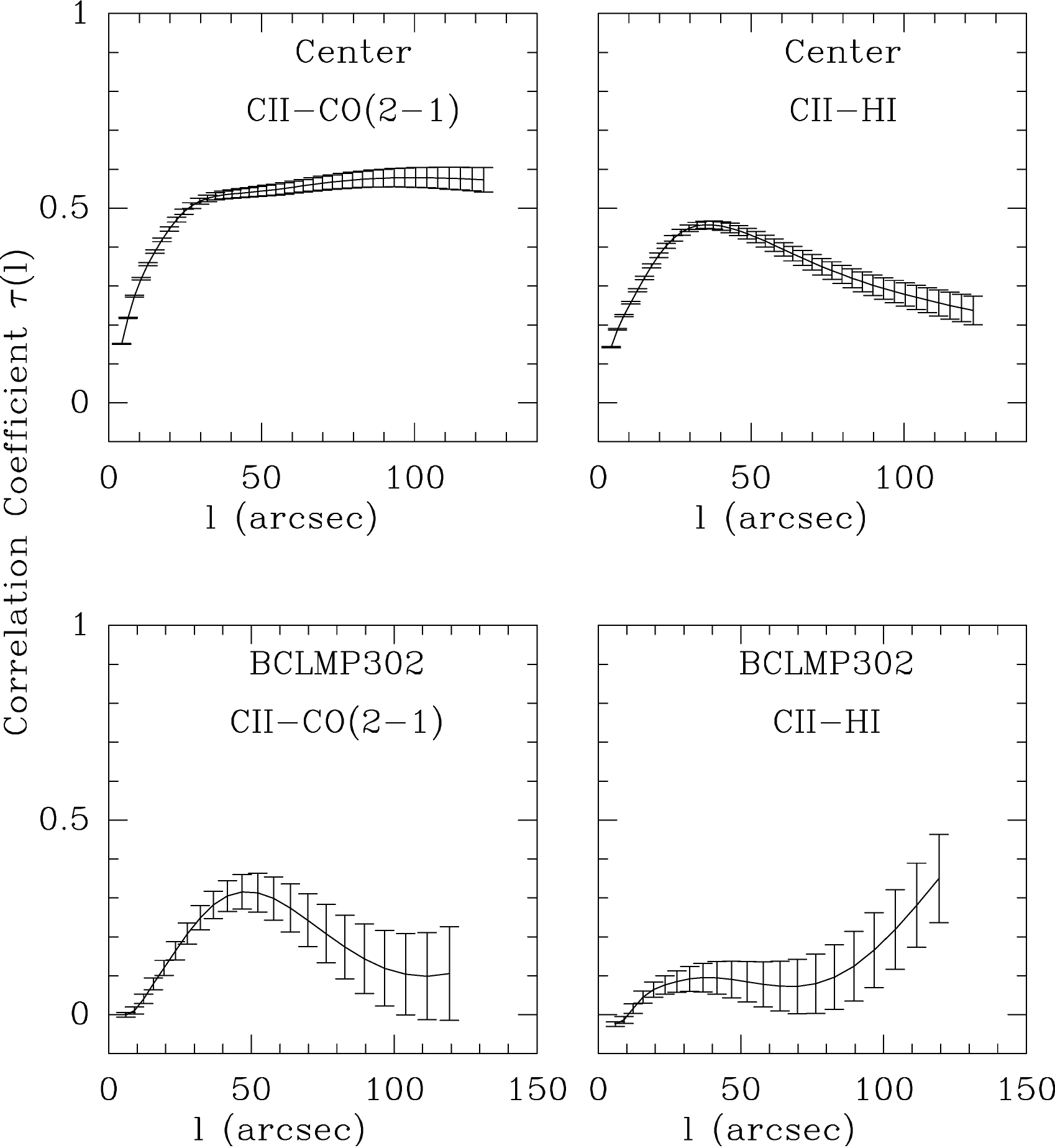}
\caption{Correlation coefficient as a function of scale for the
integrated intensity maps presented in Fig.\,\ref{fig_pacscomp}. 
\label{fig_crosscorr}}
\end{figure}

The correlation coefficients presented above measure the correspondence
between the structures in the maps, independent of their size, averaged
over the entire map.  For a scale-dependent investigation that only
compares structures with a particular size, we measure the
correspondence on a scale-by-scale basis.
For this analysis, we use the wavelet-based weighted cross-correlation
(WWCC) tool developed by \citet{arshakian2015} to study the degree of
correlation of structures seen in a pair of maps as a function of their
spatial scale. The method first filters the maps to be compared with a
wavelet of a characteristic size so that only structures of that size
remain in the maps, and in a second step computes the correlation
between the filtered maps.  With this approach, only structures of the
same scale are compared.  Use of different wavelet sizes results in a
spectrum of correlation coefficients. In case of systematic shifts of
characteristic structures between the maps, the tool can also measure
their mutual displacement.  Figure\,\ref{fig_crosscorr} shows the
correlation coefficient as a function of scale for the \CII--CO(2--1)
and \CII--\HI\ maps for both the central and the BCLMP302 regions.

In the center of M\,33, we find an increasing correlation both between
\CII\ and  CO(2--1) and between \CII\ and \HI{} at scales up to about
30\arcsec\ (Fig.\,\ref{fig_crosscorr}). As seen in the maps, individual
smaller structures are typically displaced between the different maps,
but the main emission peaks, having a size of about 30\arcsec{}, agree
in their rough location.  When considering larger scale structures, the
relative contribution of those peaks matters as they merge.  In \HI\ the
strongest emission is located in the western peak and the overall
emission is very extended while the \CII{} and CO(2--1) emission is
highly concentrated at the eastern peak.  This leads to constant and
high correlation between \CII{} and CO(2--1) at large scales and a
corresponding decreasing correlation with \HI. 

In BCLMP302, we even find a small anticorrelation between the species at
the smallest scales. The maps already show that the location of the
peaks is rather disjunct.  When comparing \CII\ and CO(2--1), the
correlation increases again up to scales of about 45\arcsec{}, the size
of the main emission feature seen in both species. The correlation
coefficient, however, is less than 30\% as the internal structure of
both peaks is very different and there is also a systematic offset. When
looking at larger scales, the correlation drops here because the two
CO(2--1) emission features in the north start to merge so that the main
structure is shifted to the north relative to the centrally peaked main
emission seen in \CII{}. When comparing \CII{} with \HI{}, the situation
is different. The coefficient is less than 10\% for all structures that
can be identified by eye in Fig.\,\ref{fig_pacscomp}, but at the largest 
scales the individual clumps in the overall extended \HI{} emission
start to merge into a centrally peaked structure that appears more
similar to the centrally peaked \CII{} map. 

Overall, we find that for both regions, the global correlation
coefficients between \CII, CO(2--1) and \HI\ maps, derived considering
structures of all sizes, are larger than even the peak of the
correlation coefficient seen on a scale-by-scale basis.  We also see
that a good correlation measured at large scales is not at all
necessarily related to a good correlation at smaller scales. Thus,
observations with insufficient spatial resolution can easily be
misleading. The possible implications of these results are discussed in
Sect.\,\ref{sec_conclusion}.

\subsection{Correlation plots\label{sec_corrplot}}

\begin{figure}[h]
\centering
\includegraphics[width=0.50\textwidth,angle=0]{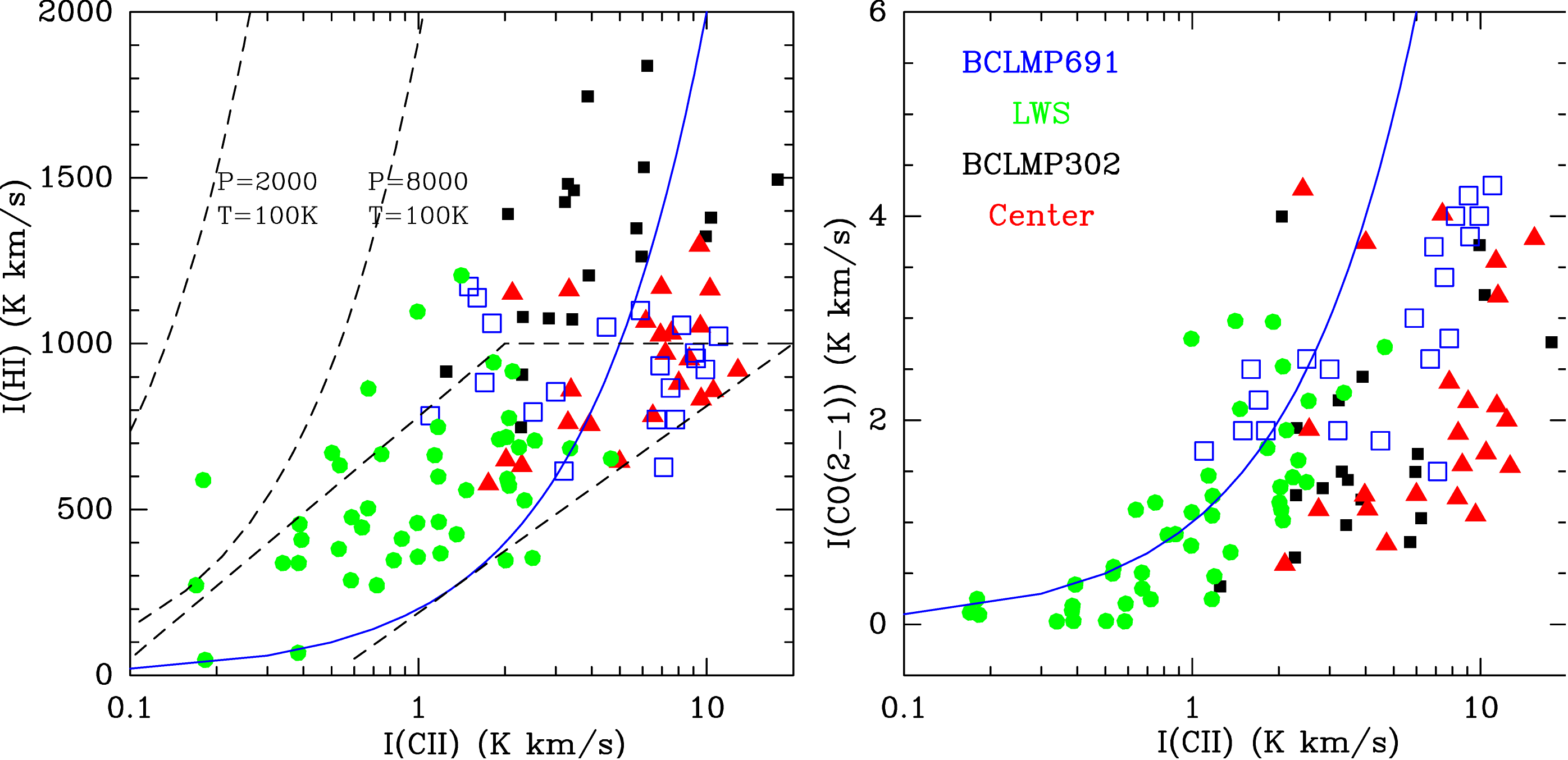}
\caption{Correlation of integrated intensities of \HI\ ({\em left}) and
CO(2--1) ({\em right}) with \CII\ integrated intensities for BCLMP302
(black squares), central region (red triangles), and BCLMP691 \citep[blue
empty squares;][]{braine2012}  and for positions along the major axis of
M\,33 measured with ISO/LWS at 280\,pc resolution
\citep[green filled circles;][]
{kramer2013}. For the central and BCLMP302 region, \CII, CO(2--1) and \HI\ 
are integrated only over the range of velocities in which \CII\ is
detected at the individual positions.
The dashed curved lines
in the left panel show the predicted $I$(\CII) for low-density diffuse
atomic \HI\ clouds as a function of $I$(\HI), corresponding to the
pressures $P$ = 2000 and 8000\,K\,\cmcub\ for a kinetic temperature of
100\,K. 
The box in dashed lines shows the region 
occupied by the \CII\ emitting \HI\ clouds  analyzed by the GOT C$^+$ project
\citep{wlanger2014}. Both panels show  lines (blue continuous) of
proportionality to guide the eye.
\label{fig_corrplot}}
\end{figure}

The velocity-resolved \CII\ spectra observed with HIFI provide us with
the unique opportunity of identifying the velocity range over which
\CII\ is emitted. Consequently, it is possible to compare the intensities
of CO(2--1) and \HI\ integrated over the same velocity range where \CII\
is detected.  Figure\,\ref{fig_corrplot} shows a comparison of the \HI\
and CO(2--1) intensities as a function of the \CII\ intensities at all
positions observed with HIFI in the center and the BCLMP302 regions.
All the integrated intensities were calculated over the velocity range
($\upsilon_{\rm min}$ to $\upsilon_{\rm max}$ in
Tables\,\ref{tab_nucfits} and \ref{tab_bclmp302fits}) in which \CII\
emission is detected.  For comparison, we have overplotted the \CII, CO,
and \HI\ intensities measured with HIFI in the BCLMP691 region
\citep{braine2012}, as well as the results of the analysis of the
ISO-LWS data taken along the major axis of M\,33 \citep{kramer2013}.
Figure\,\ref{fig_corrplot} also shows the region (dashed parallelogram)
in the \CII--\HI\ correlation graph occupied by the clouds detected in
the Milky Way by the GOT C$^+$ survey performed with Herschel/HIFI
\citep{wlanger2014}.

The \CII\ intensities measured by the ISO-LWS typically have values
lower than those measured using HIFI, possibly because of the
significantly larger size of the ISO beam (70\arcsec). 
The points belonging to the central and the BCLMP302 regions occupy
separate areas in the \CII-\HI\ correlation plot. For the same values of
\CII\ intensities, the BCLMP302 positions show higher $I$(\HI) as
compared to the positions in the central region.  This kind of a
segregation is not seen in the \CII-CO(2--1) plot where the points from
the two regions overlap significantly with each other.  The points
corresponding to BCLMP691 occupy essentially the same region in the
\CII--\HI\ correlation plot as the M\,33 center, although the \HI\
intensity from BCLMP691 remains almost constant.   For the same values
of $I$(\CII), $I$(CO(2--1)) tends to be higher in BCLMP691 than in the
other regions.  The Galactic clouds detected by the GOT \cplus\ survey
typically have similar $I$(\CII) as the M\,33 clouds but somewhat lower
$I$(\HI).  Comparison of M\,33 and Galactic molecular clouds is strongly
affected by the different spatial scales probed by the HIFI
observations.  Furthermore, for the Galactic observations within GOT
C$^+$ the beam size of \HI\ observations was larger than the \CII\ beam. 

Table\,\ref{tab_correlate} presents the linear correlation coefficients  for
the \CII--\HI\ and the \CII--CO(2--1) scatterplots
(Fig.\,\ref{fig_corrplot}) for the different datasets. We find that for
the BCLMP302, BCLMP691, and ISO-LWS data along the major axis of
M\,33, the \CII\ intensities have a correlation coefficient $>65$\% with
CO(2--1), whereas in the center the correlation is somewhat poor.  The
correlation between \CII--\HI\ also varies from region to region, with
\HI\ in BCLMP691 practically showing no correlation with \CII.  For both
the central and BCLMP302 region, the correlation coefficient for
\CII--CO(2--1) and \CII--\HI\ derived with the line velocity information
appears to contradict the correlation coefficient estimated from the
PACS observations of those region (Sec.\,\ref{sec_pacs}).

\begin{table}[h]
\caption{Linear correlation coefficients for the scatter plots in
Fig.\,\ref{fig_corrplot}. Here $I$(CO(2--1) and $I$(\HI) were estimated
by integrating only over the velocity range over which \CII\ is detected.
\label{tab_correlate}}
\begin{center}
\begin{tabular}{ccc}
\hline
\hline
Region & \CII--CO(2--1) & \CII--\HI\\
\hline
Center & 0.33 & 0.50 \\
BCLMP302 & 0.67 & 0.61 \\
BCLMP691 & 0.70 & 0.07\\
Major Axis & 0.77 & 0.44\\
All & 0.75 & 0.66\\
\hline
\hline
\end{tabular}
\end{center}
\end{table}

Thus, for the integrated intensities derived over selected velocity
ranges (as for the center, the BCLMP302, and BCLMP691 regions) the
\CII-CO(2--1) correlation varies drastically with location,
with M\,33's outer arm regions showing higher correlations.
The correlation between \CII\ and \HI\ is generally worse
than the \CII--CO(2--1) correlation, and also varies greatly between the
different regions.

\section{Results: position-velocity diagrams}

\subsection{Central region of M\,33}

\begin{figure*}[h]
\centering
\includegraphics[width=0.80\textwidth,angle=0]{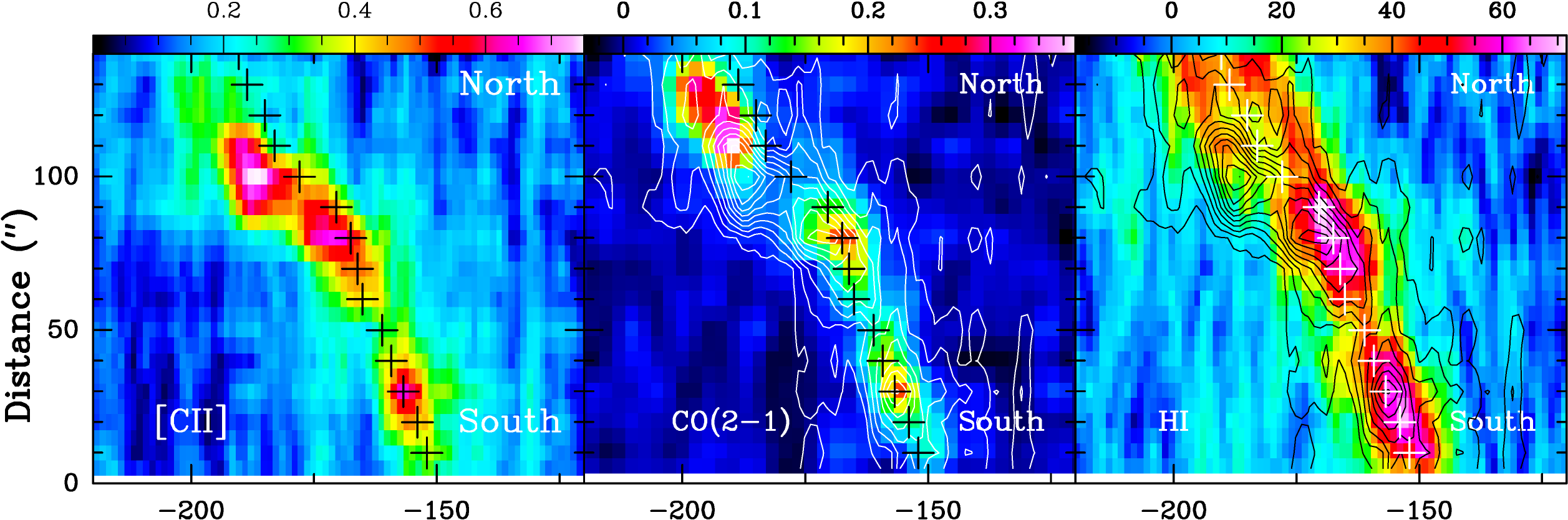}
\includegraphics[width=0.80\textwidth,angle=0]{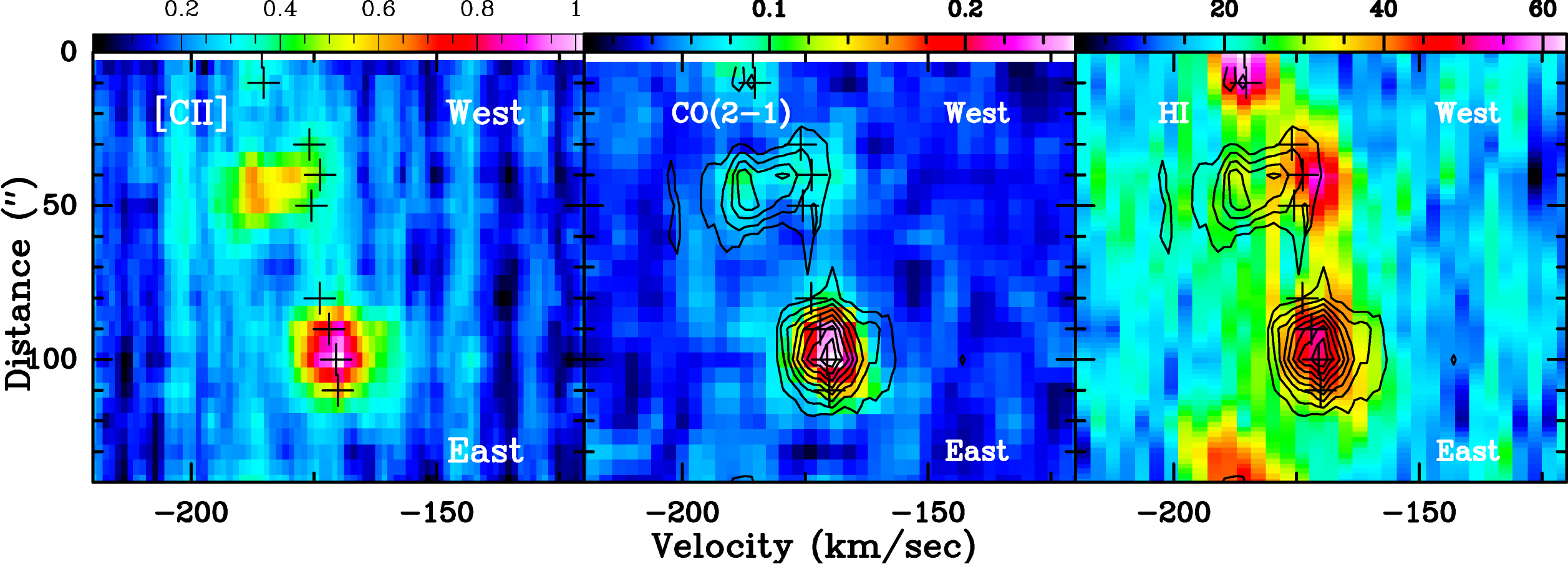}
\caption{Position-velocity diagrams for \CII, CO(2--1) and \HI\ along
the north-south ({\em top}) and east-west ({\em bottom}) directions in
the central region of M\,33. The contours in the middle and right panels
correspond to \CII\ emission and range from 10\% to 100\% of the peak
values of 0.6 and 0.8\,K in steps of 10\% for the NS and EW cuts,
respectively. The color bar on top of each panel shows the color scale of
the corresponding image in units of K. The crosses in all panels
denote the nominal velocities due to galactic rotation obtained from \HI\
spectra (see text). 
\label{fig_pvdiag_nuc}}
\end{figure*}

Figure~\ref{fig_pvdiag_nuc} shows the position-velocity (PV) diagrams
along cuts oriented in the NS and EW directions for \CII, CO(2--1) and
\HI\ of the region close to the center of M\,33. For all three tracers,
the velocity gradients observed toward the center of M\,33 are
dominated by the rotation of the galaxy. We first estimate the
differences in velocities observed in the three tracers beyond the
effect of rotation of the galaxy. For this, we have derived the velocity
due to the galactic rotation by fitting a single-component Gaussian to
\HI\ spectra smoothed to a resolution of 10\,\kms\ at individual
positions on the cut. The fitted \HI\ velocities are indicated with
crosses in  Fig.\,\ref{fig_pvdiag_nuc}.  The velocities estimated from
the \HI\ spectra along both directions are consistent with the galactic
rotation curve derived by \citet[cf.][]{corbelli2007} using \Halpha\
data.  The velocity due to rotation leads to gradients of 40\,\kms\ and
25\,\kms\ over a length of 140\arcsec\ in the NS and EW directions,
respectively.

For the NS cut, we find that the velocities of the peak emission of both
\CII\ and CO(2--1) agree reasonably well with the centroid velocity of
the \HI\ emission up to a length of 70\arcsec\ starting from the south
(Fig.~\ref{fig_pvdiag_nuc}, top).  Further north along the NS cut, (i)
the CO(2--1) peak is shifted toward more northern positions relative to
\CII, and (ii) \HI\ emission shows double emission peaks around
100\arcsec, of which the lower velocity peak is followed by \CII\ and
CO(2--1).  Along the EW direction, \HI\ shows four emission peaks, out
of which only one around 100\arcsec\ from west matches the \CII\
and CO(2--1) peaks.  The \CII\ emission peak at 50\arcsec\ from west
occurs at 16\,\kms\ lower velocity compared to \HI, while CO(2--1),
though not bright at this position, is closer in velocity to \CII. The
redward shift in \HI\ velocity relative to the velocities of \CII\ and
CO(2--1) in the central region of M\,33 could be indicative of local
dynamics, which leads to differential motion between the diffuse atomic
and molecular (or PDR) phases of the ISM. This also implies that toward
the center of the galaxy the \CII\ emitting gas is better associated
with the CO emitting PDRs than the \HI\ emitting more diffuse atomic
gas.


\subsection{BCLMP302 region}

\begin{figure*}[h]
\centering
\includegraphics[width=0.82\textwidth,angle=0]{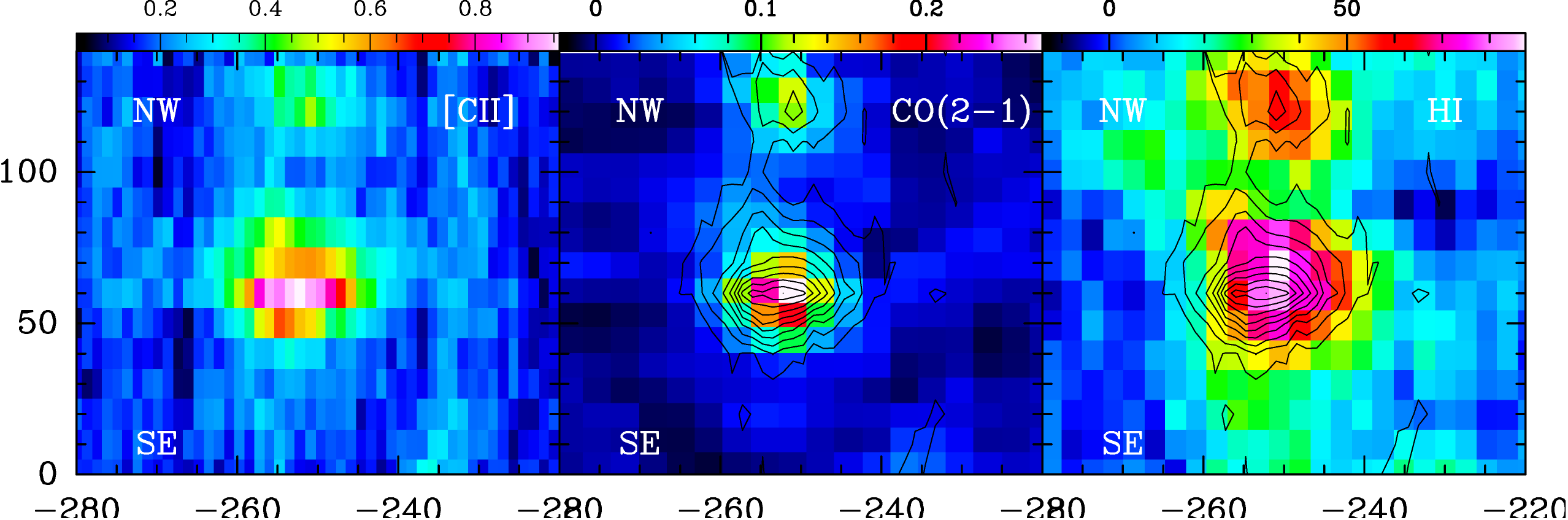}
\includegraphics[width=0.80\textwidth,angle=0]{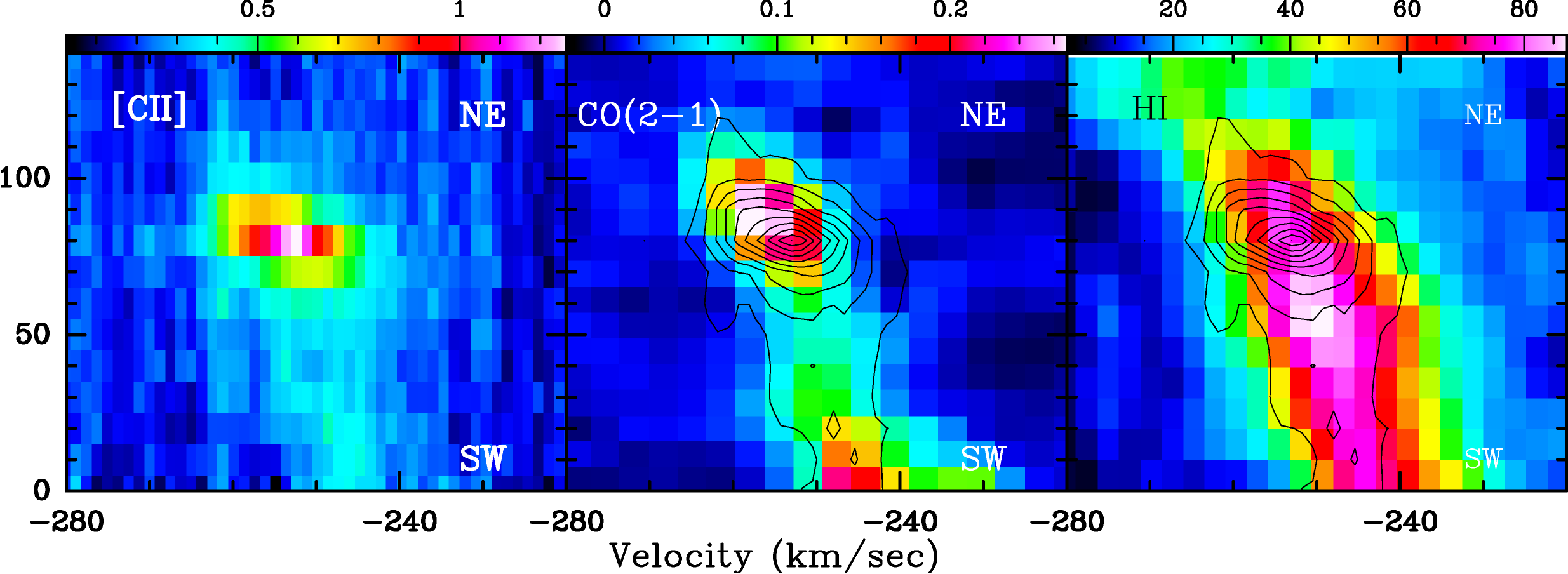}
\caption{Position-velocity diagrams for \CII, CO(2--1) and \HI\ along
the SE-NW ({\em top}) and SW-NE ({\em bottom}) directions for BCLMP302. The
contours in the middle and right panels correspond to \CII\ emission and
range from 20\% to 100\% of the peak value of 1.1\,K  and
0.8\,K in steps of 10\% for SE-NW and SW-NE cuts, respectively. The color
bar on top of each panel shows the color scale of the corresponding image 
in units of K.
\label{fig_pvdiags2}}
\end{figure*}

Similar to the central region, the \CII\ spectra in the two cuts along
the SE-NW and SW-NE directions crossing BCLMP302 show broader line
widths than the CO(2-1) lines, while the \HI\ emission is only slightly
broader than the \CII\ line (Fig.~\ref{fig_pvdiags2}).  Along the
SE-NW cut all three tracers match well both velocity-wise and spatially.
Along the SW-NE cut in the BCLMP302 region: (i) the velocities of the
\CII\ and \HI\ emitting clouds appear to match each other, while the
CO(2--1) emission shows a redward shift by 10\,\kms, and (ii) although
the \CII\ and CO(2--1) emission peaks match spatially, the \HI\ emission
peak is shifted by 10--20\arcsec\ to the SW.  

While there is an overall similarity of the \CII, CO(2--1) and \HI\
position-velocity diagrams, both in the center and BCLMP302, in some
locations strong \CII\ is detected at velocities significantly shifted
relative to both the CO(2--1) and \HI\ emission.

\section{Excess \CII\ emission seen in the spectra 
\label{sec_specfit}}

\begin{figure*}[h]
\centering
\includegraphics[width=0.80\textwidth,angle=0]{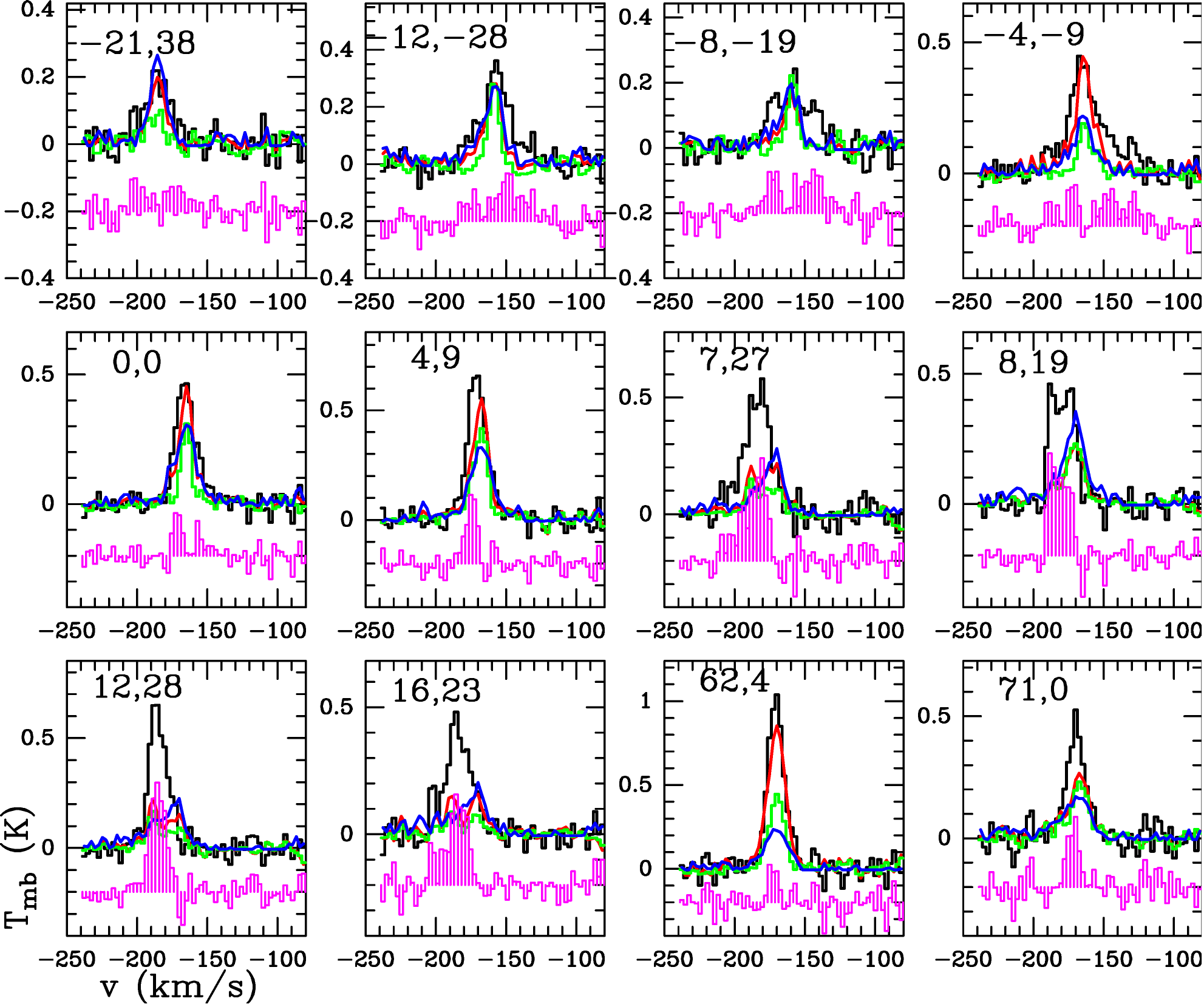}
\caption{\CII\ (black), CO(2--1) (green), and \HI\ (blue) spectra at
selected positions in the center of M\,33. The \HI\ spectrum has been
scaled by a factor of 0.005 for easier comparison. The fit to the 
observed \CII\ spectrum, as obtained from the method described in text,
is shown in red and the residual spectrum (vertically offset) is shown 
in magenta.
\label{fig_nucspec}}
\end{figure*}

\begin{figure*}[h]
\centering
\includegraphics[width=0.80\textwidth,angle=0]{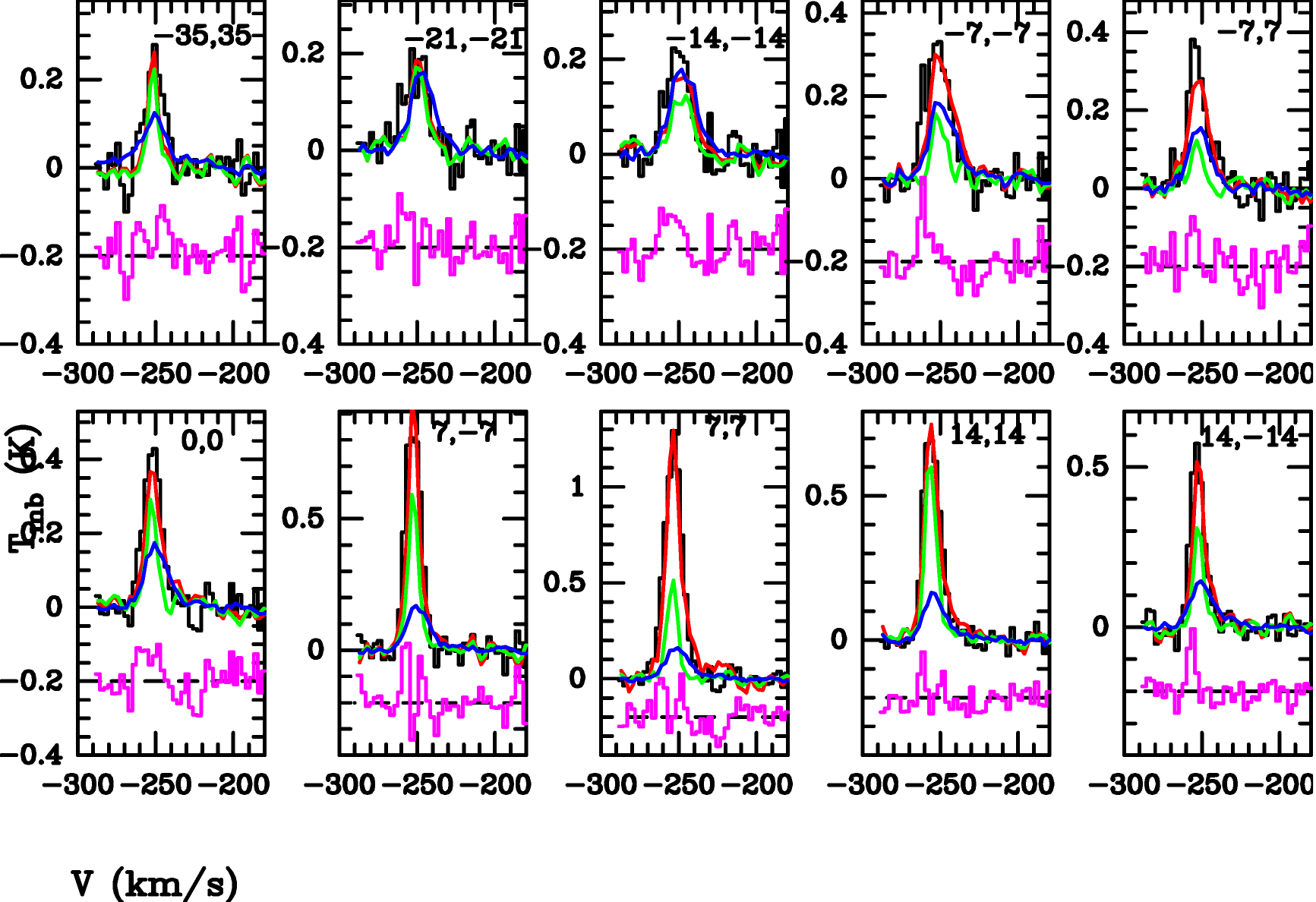}
\caption{\CII\ (black), CO(2--1) (green), and \HI\ (blue) spectra at
selected positions in the BCLMP302 region. The \HI\ spectrum has been
scaled by a factor of 0.002 for easier comparison. The fit to the 
observed \CII\ spectrum, as obtained from the method described in text,
is shown in red and the residual spectrum (vertically offset) is shown in magenta.
\label{fig_bclmp302spec}}
\end{figure*}

Figures\,\ref{fig_nucspec} and \ref{fig_bclmp302spec} show the \CII,
CO(2--1) and \HI\ spectra at selected positions in the central and
BCLMP302 regions of M\,33. The positions in both regions were chosen
based on the S/N ratio of the \CII\ spectra. In the central region, the
\CII\ spectra show widths intermediate between the linewidths of \HI\
and CO(2--1). Given the higher propensity of CO(2--1) to become
optically thick it is understandable if both \CII\ and \HI\ emission
trace larger column densities of gas \citep{henkel1997}.  As seen in the
PV diagrams, at multiple positions the \CII\ spectra show additional
features at velocities where very little or no CO(2--1) and/or \HI\
emission are detected. 


Ignoring the contribution of the ionized gas, the \CII\ emission may
arise from (a) HI, (b) CO-traced H$_2$, (c) H/H$_2$ not traced by either
\HI\ or CO emission. In order to estimate the fraction of \CII\ emission
arising from the component not detected in CO or \HI,
we fit the observed \CII\ spectrum at each position as a linear
combination of the CO(2--1) and \HI\ intensities. The fitting procedure
does not make any assumption about the physical heating model (PDR, GMC,
shocked ISM, XDR, CR). It maximizes the emission that could be
associated with (a) and (b), and so determines the {\em minimum} profile
associated with (c).  Thus at each position we fit the \CII\ spectrum
channel-wise as 

\begin{equation}
I ({\rm CII}) = a\,I({\rm CO})+b\,I({\rm HI}),
\end{equation}
where $a$ and $b$ are held constant for a particular spectrum so as to
obtain the best fit to the spectrum.  The fit we perform is  restricted,
and ensures that the residuals from the fit always remain positive
so that the contributions from the gas components other than those
emitting in CO(2--1) and \HI\ are always taken into account.  In order
to perform this fitting we smoothed the \CII\ and \HI\ spectra to
the resolution of the CO(2--1) spectrum, which is 2.6\,\kms.
Figures\,\ref{fig_nucspec}  and \ref{fig_bclmp302spec} also show the
fitted profile and the residual spectrum for the selected positions in
the central and the BCLMP302 regions, respectively.

At each position, we derive the fraction of \CII\ intensity arising from
the gas not detected in CO(2--1) or \HI\ by integrating the observed and
residual spectra over the entire velocity range (between $\upsilon_{\rm
min}$ and $\upsilon_{\rm max}$) in which \CII\ emission is detected
(Table\,\ref{tab_nucfits}). For all positions in the central region the
analysis outlined above suggests that 11--60\% 
of the observed \CII\ intensity is due to gas not
detected in CO(2--1) and \HI. At all positions but one, more than 20\% of
the \CII\ intensity is not explained by the detected components of
molecular and atomic gas.  We note that this process of fitting is
naturally biased toward not allowing the linear combination of CO(2--1)
and \HI\ spectra to fully fit the \CII\ intensity.  Additionally, the
fitting procedure cannot break the degeneracy between the contributions
from the gas emitting CO(2--1) and \HI\ when the two profiles match
reasonably well, barring the differences in absolute intensities.  Since
we have insufficient information to distinguish this, hence, the
designation minimum for the contribution of the CO-dark molecular gas. The
existence of untraced H/H$_2$ gas makes the a priori use of any model,
PDR or otherwise, to determine physical conditions somewhat speculative.

\begin{table*}[h]
\caption{Results of analysis of \CII, CO(2--1) and \HI\ spectra for
selected positions in the central region. The quantities $I$(\CII),
$I$(CO(2--1) and $N$(\HI) = 1.82$\times 10^{18}$ $I$(\HI) (with $I$(\HI) in K\,\kms) are estimated 
over the velocity range $\upsilon_{\rm min}$ to $\upsilon_{\rm max}$.  $I_{\rm res}$(\CII), 
$a$, and $b$ are obtained from fitting the \CII\ spectrum as a combination of CO(2--1) 
and \HI\ spectra, as described in Sect. 6.
\label{tab_nucfits}}
\begin{tabular}{cccccccccc}
\hline
Offset & $\upsilon_{\rm min}$ & $\upsilon_{\rm max}$ & $I$(\CII) & 
 $I$(CO(2--1)&  $N$(\HI)& $I_{\rm res}$(\CII) &  $\frac{I_{\rm res}(CII)}{I(CII)}$ &$a$ & $b$ \\
(\arcsec,\arcsec) &\kms &\kms & K\,\kms &  K\,\kms & 10$^{21}$cm$^{-2}$ & K\,\kms &  &  & \\
\hline
\hline
(-21, 38)  &  -210 &   -160   &   4.64$\pm$0.81 &  1.3 &   1.4  &  1.78 &   0.38  &   0.4  &  3.1e-3   \\
(-12,-28)  &  -181 &   -129   &   8.25$\pm$0.61 &  2.0 &   1.9  &  4.04 &   0.49  &   0.4  &  3.4e-3   \\
( -8,-19)  &  -186 &   -135   &   6.02$\pm$0.60 &  2.1 &   1.2  &  3.60 &   0.60  &   0.3  &  2.8e-3   \\
( -4, -9)  &  -198 &   -121   &  12.64$\pm$0.79 &  2.5 &   1.6  &  4.81 &   0.38  &   0.9  &  6.3e-3   \\
(  0,  0)  &  -189 &   -140   &   8.36$\pm$0.60 &  3.1 &   2.1  &  1.90 &   0.23  &   0.8  &  3.5e-3   \\
(  4,  9)  &  -192 &   -148   &  11.36$\pm$0.64 &  5.9 &   2.4  &  3.03 &   0.27  &   1.1  &  1.4e-3   \\
(  7, 27)  &  -220 &   -155   &  12.25$\pm$1.00 &  3.3 &   2.1  &  6.51 &   0.53  &   1.1  &  1.6e-3   \\
(  8, 19)  &  -197 &   -164   &   9.04$\pm$0.50 &  3.8 &   2.0  &  5.36 &   0.59  &   0.8  &  6.7e-4   \\
( 12, 28)  &  -206 &   -165   &  10.47$\pm$0.63 &  2.7 &   1.7  &  6.31 &   0.60  &   1.2  &  1.0e-3   \\
( 16, 23)  &  -213 &   -163   &   9.64$\pm$0.83 &  1.7 &   1.6  &  6.02 &   0.62  &   1.6  &  9.4e-4   \\
( 62,  4)  &  -193 &   -135   &  15.43$\pm$1.08 &  6.2 &   1.7  &  1.68 &   0.11  &   1.3  &  5.9e-3   \\
( 71,  0)  &  -190 &   -145   &   7.77$\pm$0.85 &  3.9 &   1.4  &  2.88 &   0.37  &   0.9  &  1.8e-3   \\ 
\hline
\hline
\end{tabular}
\end{table*}

\begin{table*}[h]
\caption{Same as Table\,\ref{tab_nucfits}, but for selected positions in 
the BCLMP302 region.
\label{tab_bclmp302fits}}
\begin{tabular}{cccccccccc}
\hline
Offset & $\upsilon_{\rm min}$ & $\upsilon_{\rm max}$ & $I$(\CII) & 
 $I$(CO(2--1)&  $N$(\HI)& $I_{\rm res}$(\CII) &  $\frac{I_{\rm res}(CII)}{I(CII)}$ &$a$ & $b$ \\
(\arcsec,\arcsec) &\kms &\kms & K\,\kms &  K\,\kms & 10$^{21}$cm$^{-2}$ & K\,\kms &  &  & \\
\hline
\hline
(-35, 35) & -268 & -237 &  3.42$\pm$0.45 &  1.6  &  2.0 &  1.09 &  0.32  &  1.0  &   6.7e-4 \\
(-21,-21) & -276 & -232 &  3.49$\pm$0.50 &  2.3  &  2.7 &  0.81 &  0.23  &  1.0  &   2.3e-4 \\
(-14,-14) & -270 & -233 &  3.90$\pm$0.48 &  2.0  &  3.2 &  0.70 &  0.18  &  0.7  &   1.0e-3 \\
( -7, -7) & -272 & -227 &  6.12$\pm$0.57 &  1.7  &  3.4 &  0.44 &  0.07  &  0.3  &   2.8e-3 \\
(  0,  0) & -268 & -228 &  6.09$\pm$0.39 &  2.7  &  2.8 &  0.66 &  0.11  &  0.8  &   2.2e-3 \\
(  7,  7) & -274 & -227 & 17.61$\pm$0.55 &  4.5  &  2.7 &  1.12 &  0.06  &  1.7  &   5.9e-3 \\
( 14, 14) & -269 & -237 &  9.93$\pm$0.41 &  6.1  &  2.4 &  0.50 &  0.05  &  0.8  &   3.7e-3 \\
( 14,-14) & -274 & -233 &  5.93$\pm$0.45 &  2.5  &  2.3 &  0.44 &  0.07  &  1.3  &   1.8e-3 \\
\hline
\hline
\end{tabular}
\end{table*}

Figure\,\ref{fig_bclmp302spec} shows the  observed \CII, CO(2--1) and
\HI\ spectra (all smoothed to 2.6\,\kms) for ten selected positions in
the BCLMP302 region,  along with the fit to the spectrum at each
position derived using the described method and the residual spectrum.
We find that (as mentioned before) the \CII\ spectrum has a line width
intermediate between the CO(2--1) and \HI\ spectra and the peak velocity
of \CII\ is slightly shifted. However, unlike the \CII\ spectrum from
the central region, in BCLMP302 the spectral profiles of \CII, CO(2--1)
and \HI\ look similar with very little \CII\ emission detected at
velocities not detected in the other two tracers. This is different from
the conclusions derived on the basis of the integrated intensity maps of
the region (Fig.\,\ref{fig_pacscomp}), which show very little similarity
in the emission features due to the three tracers.  We performed fits to
the spectra at all positions in the BCLMP302 region following the
procedure presented above and find that the contribution of the gas not
traced by either CO(2--1) and \HI\ ranges between 10-55\%, with typical
values around 20\% (Table\,\ref{tab_bclmp302fits}).

\section{Emission of \CII\ of ionized gas}

In order to investigate the origin of \CII\ intensity not directly
assignable to the observed CO(2--1) and \HI\ emission, we estimate the
possible contribution of \HII\ regions in the center and in BCLMP302.
Among the \HII\ regions observed by \citet{israel1974}, the \HII\
regions \# 35, 37, 38, and 42 are located in the central region and 
\#53 corresponds to BCLMP302.

The relation between the \CII\ intensity (in K\,\kms) and the column
density, $N$(\cplus) is \citep{crawford1985,pineda2013} 

\begin{equation}
I({\mathrm CII}) = N({\mathrm C^+})\left[3.05\times
10^{15}\left(1+0.5\left(1+\frac{A_{ul}}{qn}\right)\exp^{91.2/kT_{\rm
kin}}\right)\right]^{-1},
\end{equation}
where $A_{\rm ul} = 2.3 \times 10^6$\,s$^{-1}$ is the Einstein
spontaneous decay rate and $q$ is the collisional de-excitation
rate coefficient at a kinetic temperature of $T_{\rm kin}$.  We assume
$n=100$\,\cmcub\ and $T_{\rm kin}$ = 10000\,K (typical for \HII\
regions). For collisions between \cplus\ and electrons the de-excitation 
coefficient $q$ =
1.4$\times 10^{-6}$T$_{\rm kin}^{-0.37}$ = 4$\times
10^{-8}$\,s$^{-1}$\,\cmcub\ \citep{goldsmith2012}. Thus we get $I$(\CII)
[K\,\kms]= $N$(\cplus)/5.5$\times10^{15}$. We use the emission measure
(EM = $n_e^2 D$, where $D$ is the size of the source) observed by
\citet{israel1974} to determine $N$(H) (assuming H to be completely
ionized, $n_e$ = $n$(H$^+$) and $N$(H$^+$)=$N$(H)=$n$(H)$D$) for
$n$=100\,\cmcub.  Since the metallicity of M\,33 as determined by
\citet{magrini2010} does not vary significantly within the inner
2.5\,kpc, we use C/H = 8$\times 10^{-5}$ (assuming no depletion, since to
a good approximation there are no grains in \HII\ regions)
to estimate $N_{\rm ion}$(\cplus) from $N$(H) for all \HII\
regions.

\begin{table}[h]
\caption{Estimate of \CII\ emission from ionized gas.
\label{tab_cplusion}}
\begin{tabular}{lccccc}
\hline
\# & ($\Delta\alpha$,$\Delta\delta$) & EM & $N$(H) & $N_{\rm ion}$(\cplus) & $I_{\rm ion}$(CII) \\
&  & 10$^3$ & 10$^{20}$ & 10$^{16}$ &  \\
& (\arcsec,\arcsec) & pc\,cm$^{-6}$ & \cmsq & \cmsq & K\,\kms \\
\hline
\hline
{\tiny Center} &&&&&\\
35 & (-15,-37) & 3.0 & 0.93& 0.74 & 1.3$^a$ \\
37 & (0,0) & 19.1 & 5.9 & 4.7 &
2.1$^{b}$ \\
38 & (16,23) & 5.2 & 1.6 & 1.3 & 1.7 \\
42 & (62,4) & 3.6 & 1.1 & 0.89 & 1.6 \\
{\tiny BCLMP302} &&&&&\\
53 & (7,7) & 3.0 & 0.93 & 0.74 & 1.3 \\
\hline
\end{tabular}

$^a$ $I$(\rm CII) at (-15,-37) is 11.4\,K\,\kms.\\
$^b$ Using a dilution factor of 0.20 to account for the smaller size of
the \HII\ region (5\farcs4) compared to the 12\arcsec\ resolution of
\CII\ data.
\end{table}

Table\,\ref{tab_cplusion} presents the parameters for the four \HII\
regions as obtained from \citet{israel1974}, along with estimates of the
\CII\ intensity originating from the ionized gas.  We emphasize that only the
\HII\ region \#37 is smaller (5\farcs4) than the HIFI beamsize and the
$I$(\CII) estimated to arise from the ionized gas had to be corrected
for the beam dilution.  For the positions in the central region: at (-15,-37),
(0,0), (16,23), and (62,4) the ionized gas contributes 11\%, 25\%, 18\%,
and 10\% respectively, to the total \CII\ emission.   In BCLMP302, we find
that while 14\% of \CII\ emission is unexplained at (7,7), only 7\% can
arise from the ionized gas.  \citet{mookerjea2011} had used the measured
\NII\ (122\,\micron) intensities to estimate a contribution of 20--30\%
by the ionized gas in BCLMP302. Both of these estimates depend
crucially on the value assumed for $n_{\rm e}$. For smaller values of
$n_{\rm e}$ the contribution of the ionized gas to the \CII\ intensity
is larger, as less Carbon is driven to higher ionization
states \citep{abel2006}.   In the present estimate, as well as in
\citet{mookerjea2011}, a density of $n_{\rm e}$ = 100\,\cmcub\ has been
used and the contribution of ionized gas to the overall \CII\ emission
is found to be $<$25\% in the two regions of M\,33. 

\section{Properties of the neutral atomic gas emitting \CII}

We use the results of the fitting procedure to estimate the fraction
of \CII\ emission arising from the molecular gas traced by CO(2--1)
and atomic gas traced by \HI\ (Tables\,\ref{tab_nuccomp} and
\ref{tab_b302comp}). 
\citet{kramer2013} had estimated that for positions along the major axis
of M\,33, PDRs contribute a constant \HI\ column density of
3.25$\times$10$^{20}$\,\cmsq, which is 15--70\% of the total \HI\ column
density they observed. They had concluded that diffuse atomic clouds
(the cold neutral medium) contribute $\sim$15\% of the observed \CII\
emission in the inner 2\,kpc of M\,33, while this contribution rises to
about 40\% in the outer disk at 6\,kpc.  Here for the central and
BCLMP302 regions the observed $N$(\HI) lies between
9.0$\times$10$^{20}$\,\cmsq\ and 3.4$\times$10$^{21}$\,\cmsq, so that
the PDR contribution to the \HI\ intensity lies at 10--30\%. The typical
\HI\ column density of diffuse atomic clouds is
$<$few$\times$10$^{20}$\,\cmsq, while that for the atomic envelopes of
dense molecular clouds lie in the range (1-7)$\times10^{20}$\,\cmsq\
\citep{wolfire2010}. This suggests that the observed \HI\ emission
either arises from several diffuse clouds or from the atomic envelopes
of dense molecular clouds, which also emit in \CII.  This is consistent
with \citet{kramer2013} and the results of the Galactic \CII\ survey by
\citet{wlanger2014}, which had concluded that the \HI\ emission is much
brighter than what is expected from a single diffuse atomic cloud.

It is not possible to separate out the contribution of the diffuse and
dense (PDR) atomic gas based on velocity information. However, since
the PDR contribution to \HI\ intensity is less than 30\% for the two
regions considered here, we use the total \HI\ intensities for the
following calculations.  We use the fraction of \CII\ intensity
estimated in the fits (Sec.\,\ref{sec_specfit}) to arise from the atomic
gas (Tables\,\ref{tab_nuccomp} and \ref{tab_b302comp}), the observed
$N$(\HI) (Tables\,\ref{tab_nucfits} and \ref{tab_bclmp302fits}) and Eq.
(2) to estimate the volume density ($n$(\HI)) of atomic gas, which could
account for the \CII\ emission.  We assume the temperature of the atomic
gas to be 100\,K and at this temperature the de-excitation coefficient
for \cplus-H collisions  is 8.1$\times10^{-10}$\,s$^{-1}$\,\cmcub\
\citep{launay1977}. For the atomic gas, we estimate that 50\% of the
carbon is locked in the dust grains so that the C/H ratio is half the
value used for the ionized gas; C/H = 4$\times 10^{-5}$.  We find that
at most positions in the center the atomic gas densities are larger than
200\,\cmcub\ and go up to 1700\,\cmcub\ at the position of the brightest
\HII\ region (Table\,\ref{tab_nuccomp}). Similarly, in the BCLMP302
region the atomic gas densities are larger than 150\,\cmcub\ and go up
to 1500\,\cmcub.  Thus the \HI\ gas contributing to the \CII\ emission
has densities significantly larger than the densities found in diffuse
clouds and are most likely to be atomic envelopes of molecular PDRs. 

\begin{table}[h]
\caption{Relative contribution of CO(2--1) and \HI\ emitting molecular
and atomic gas to \CII\ emission at positions in the central region of M\,33
$I_{\rm CO}$(CII) and $I_{\rm HI}$(CII) are the estimated integrated 
intensities of \CII\ obtained from the fits described in the text.
The difference  between the observed \CII\ intensity and the result of 
this fit is $I_{\rm res}$(CII).
\label{tab_nuccomp}}
\begin{center}
\begin{tabular}{cccccc}
\hline
\hline
Offset & $\frac{I_{\rm CO}(CII)}{I(CII)}$ & $\frac{I_{\rm HI}(CII)}{I(CII)}$ &
$\frac{I_{\rm res}(CII)}{I(CII)}$ &$n$(\HI) \\
(\arcsec,\arcsec) & \% & \% & \% & \cmcub\\
\hline
(-21, 38) &    11  &    51  &   38  &  590\\
(-12,-28) &     9  &    42  &   49  &  670\\
(-8,-19)  &     9  &    31  &   60  &  520\\
(-4,-9)   &    18  &    44  &   38  &  1700\\
( 0,0)    &    29  &    48  &   23 &  690\\
( 4,9)    &    57  &    16  &   27  &  230\\
( 7,27)   &    32  &    15  &   53  &  270\\
( 8,19)   &    33  &     8  &   59  &  100\\
(12,28)   &    31  &     9  &   60  &  160\\
(16,23)   &    28  &     9  &   63  &  150\\
(62,4)    &    53  &    36  &   11  &  1500\\
(71,0)    &    45  &    18  &   37  &  310\\
\hline
\hline
\end{tabular}
\end{center}
\end{table}

\begin{table}[h]
\caption{Same as Table\,\ref{tab_nuccomp}, but for positions in the 
BCLMP302 region. 
\label{tab_b302comp}}
\begin{tabular}{ccccc}
\hline
\hline
Offset & $\frac{I_{\rm CO}(CII)}{I(CII)}$ & $\frac{I_{\rm HI}(CII)}{I(CII)}$ &
$\frac{I_{\rm res}(CII)}{I(CII)}$ &$n$(\HI) \\
(\arcsec,\arcsec) & \% & \% & \% & \cmcub\\
\hline
(-35, 35) &    47  &   21 &  32   &   104\\
(-21,-21) &    67  &    10 &  23   &   34\\
(-14,-14) &    37  &   45 &  18   &   160\\
(-7,-7)   &     7  &   86 &   7   &   532\\
( 0,0)    &    34  &   55 &  11   &   383\\
( 7,7)    &    43 &   51 &   6   &   1540\\
(14,14)   &    46  &   49 &   5   &   740\\
(14,-14)  &    54  &   39 &   7   &   316 \\
\hline
\hline
\end{tabular}
\end{table}

\section{Estimate of carbon abundances at selected positions}

We next check the consistency of the derived column densities of \cplus,
C$^0$, and CO with the abundance of gaseous carbon relative to hydrogen.
To accomplish this, we select two positions, one each in the two regions
presented here, for which we also have low-$J$ CO ($^{13}$CO)
observations at an angular resolution similar to the \CII\ observations. 

For GMC1 \citep{gratier2012}, a molecular cloud in the central region
(offset 62\arcsec,4\arcsec), an LTE analysis of the CO (and $^{13}$CO)
intensities of the $J$=1--0 and $J$=2--1 was performed by
\citet{buchbender2013}. For the BCLMP302 region, \citet {mookerjea2011}
have presented the CO($^{13}$CO) intensities based on IRAM 30m
observations at a position almost coincident with the offset
(7\arcsec,7\arcsec). The fraction of \CII\ emission contributed by the
molecular, atomic, and gas unassociated with CO and \HI\ emission  in
GMC1 is 53\%, 36\%, and 11\%. In BCLMP302 it is 43\%, 51\%, and 6\%. The
\HII\ region contributes 11\% and 7\% of the total \CII\ emission in
GMC1 and BCLMP302, respectively.  The density of atomic gas required to
explain the fraction of \CII\ emission arising from it is 1500\,\cmcub\
for both positions.

For both positions, we estimate the CO column densities from the
$^{13}$CO(1--0) intensities assuming LTE and optically thin emission

\begin{equation}
N =
\frac{3h}{8\pi^3 S\mu^2}\frac{Z}{g_J}\frac{\exp{\left(\frac{h\nu}{kT_{\rm
ex}}\right)}}{\left[\exp{\left(\frac{h\nu}{kT_{\rm
ex}}\right)-1}\right]}\left[J_\nu(T_{\rm ex})-J_\nu(T_{\rm
BG})\right]^{-1}\int{T_{\rm mb}d\upsilon},
\end{equation}
where $\mu$ = 0.11 Debye is the dipole moment for CO, $S=J/(2J+1)$ is
the line strength, $Z$ = $\sum\limits_{J=0}^{\infty}
g_J\exp\left(-\frac{E_J}{kT}\right)$ is the partition function, $E_J$ =
$hBJ(J+1)$, $g_J = (2J+1)$, $J_\nu(T) =
\frac{h\nu}{k}\frac{1}{\exp^{{h\nu}/kT}-1}$.  We assume the excitation
temperatures (T$_{\rm ex}$) to be the same as the dust temperatures at
the selected positions and use the values given by
\citet{buchbender2013}. For GMC1 we use the $I$(CO(1--0)) from Table 2 of
\citet{buchbender2013}, $I$(CO(2--1)) from this work and for position
(7,7) in BCLMP302, we use Table 2 of \citet{mookerjea2011}.

Next for the CO(1--0) and (2--1) intensities, we use the abovementioned
excitation temperatures as kinetic temperatures and the estimated LTE
column densities to estimate the local volume densities ($n$) based on
an LVG analysis using RADEX \citep{vandertak2007}. The values of $n$ at
GMC1 and BCLMP302 are 3.3$\times 10 ^3$\,\cmcub\ and 9$\times
10^3$\,\cmcub, respectively.  For these densities ($n$), we assume a kinetic
temperature of 91\,K and use for the \cplus-H$_2$ collisions
de-excitation coefficients given by $q = \left(4.9+0.22\frac{T_{\rm
kin}}{100K}\right)\times 10^{-10}$ \citep[Eq. 3 in][]{wiesenfeld2014} to
estimate $N$(\cplus) at the selected positions using Eq. 2.
Table\,\ref{tab_colden1} presents the parameters and estimated \cplus\
and CO column densities for the two positions in the center and 
BCLMP302 region. We find that the two positions have very similar values
of $N$(\cplus), although GMC1 has a $N$(CO), which is three times higher
than the value of the BCLMP302 position. The BCLMP302 position, on the
other hand has a 1.6 times higher atomic hydrogen column density. In low
metallicity galaxies like the SMC and LMC recent observations have shown
that \CI\ has a column density  similar to or within a factor of 4 of
the CO column density \citep[][Requena-Torres et al in
prep.]{okada2015}.  In the absence of \CI\ observations, we assume that
$N$(CO) = $N$(C$^0$) and calculate the total carbon column density, $N$(C),
at the two selected positions as $N$(\cplus)+$N$(C$^0$)+$N$(CO).

\citet{braine2010} derived a  H$_2$ column density map for M\,33 based on
the 500\,\micron\ SPIRE map, $\beta=2$ and the 250/350\,\micron\ dust
temperature. These authors proposed that within the inner 2\,kpc of M\,33
the $N$(H$_2$) scales as 2.2$\times 10^{20} I$(CO(2--1)). For the two
selected positions we estimate $N$(H$_2$) with this scaling relation
and the measured $I$(CO(2--1)) (Table\,\ref{tab_colden1}). The total
hydrogen column density ($N_{\rm tot}$ = $N$(H)+2$N$(H$_2$)) at both
these positions results to 5.1$\times$10$^{21}$\,\cmsq. Thus
we estimate the abundance of carbon in the center and in the BCLMP302
region to be  5.6$\times 10^{-5}$ and 4.1$\times 10^{-5}$ and it is
consistent with our assumption of 50\% depletion of carbon onto the
grains.

\begin{table*}[h]
\caption{Estimate of column densities \label{tab_colden1}.}
\begin{tabular}{ccccccccccc}
\hline
\hline
Position & $I$(CO(1--0)) & $I$(CO(2--1)) & $I$($^{13}$CO(1--0)) & $T_{\rm ex}^a$ &
$N$(CO)$^b$ & $n$(H$_2$)$^c$ & $N$(\cplus)$^d$  & $N$(H$_2$)$^e$ &
$N$(\HI)$^f$\\
& K\,\kms & K\,\kms & K\,\kms & K & 10$^{16}$\cmsq & \cmcub & 10$^{17}$\cmsq &
10$^{21}$\cmsq & 10$^{21}$\cmsq\\
\hline
Center &&&&&&\\
(62,4) & 7.2 & 7.7 & 0.799 & 25 & 8.3 & 3300 & 2.0 & 1.7 & 1.7\\
\hline
BCLMP302 &&&&&&\\
(7,7) & 3.2 & 5.3 & 0.28 & 22 &  3.1 &   9000 & 1.6 & 1.2 & 2.7\\
\hline
\end{tabular}

$^a$ Equal to the $T_{\rm dust}$ as in Table D.1 of
\citet{buchbender2013}.\\
 $^b$ Calculated from $I$($^{13}$CO(1--0)), $T_{\rm ex}$ and
 ${12}$C/$^{13}$C = 60.\\
 $^c$ Estimated from LVG analysis of CO(1--0) and CO(2--1) intensities,
 with $T_{\rm dust}$ as T$_{\rm kin}$ and $N$(CO) as inputs.\\
 $^d$ Estimated using $n$ derived from $I$(CO(1--0)) and
$I$(CO(2--1)) using LVG model, $T_{\rm ex}$ = 91\,K and $I$(\CII) in
Tables\,\ref{tab_nucfits} \& \ref{tab_bclmp302fits} in Eq. 2.\\
$^e$ Estimated
using $N$(H$_2$) = 2.2$\times 10^{20}$\,$I$(CO(2--1))
\citep{braine2010}.\\
$^f$ See Tables\,\ref{tab_nucfits} and \ref{tab_bclmp302fits}.
\end{table*}

\section{Discussion 
\label{sec_disc}}

\subsection{Phases of ISM contributing to \CII\ emission}

One of the primary aims of the HerM\,33es project has been to
disentangle the contribution of the different phases of the ISM toward
\CII\ emission in M\,33. In this paper, we have used velocity-resolved
\CII\ spectra to estimate the contributions of dense PDRs (traced
by CO(2--1)), atomic gas (dense and diffuse), ionized gas, and
the CO-dark molecular gas to the observed \CII\ emission.  In the two
regions of M\,33 studied here, we find that the 
contributions of the molecular gas traced by CO and atomic gas traced
by \HI\ to the \CII\ intensity vary substantially from position to
position.  The contribution of the ionized gas to the \CII\ intensity is
estimated to be typically between 10--20\%.  For the Milky Way, the ISM
components have been shown to have approximately comparable
contributions to the \CII\ luminosities: dense PDRs (30\%), cold \HI\
(25\%), CO-dark H$_2$ (25\%), and ionized gas (20\%) \citep{pineda2014}.
Furthermore, \citet{pineda2013} concluded that the fraction of the CO-dark
molecular gas traced by \CII\ ranges from about 20\% in the metal-rich
inner Galaxy to 80\% in the metal poor outer-Galaxy. M\,33 has
approximately half the solar metallicity  and in the outskirts of the
Galactic disk the metallicity is comparable to the metallicity of M\,33.
In addition to being at a metallicity twice the value of M\,33, these
observations of the long lines of sight through the Galactic disk are
sensitive to significant contributions from the diffuse gas and, hence,
may not be directly comparable to the observations of M\,33.  Recent
SOFIA observations of regions in the SMC and LMC are probably more directly comparable to the case of M\,33. In
the star-forming regions of the SMC at a resolution of 10\,pc, the \CII\
emission is found to trace between 50--85\% of the bulk of the mass of
the molecular gas, with minor contributions from the ionized gas in the
\HII\ regions (Requena-Torres et al. in prep.). For N159 in the LMC,
\citet{okada2015} concluded that the  fraction of the \CII\ emission
that cannot be attributed to the material traced by the CO line
profiles, is around 20\%  around the CO cores and up to 50\% in the
area between the cores. In M\,33, though we do not have the spatial
resolution to distinguish the core and inter-core regions, the fraction
of \CII\ intensity tracing the CO-dark molecular gas is similar to the
value found in the LMC.  In the LMC, the ionized gas contributes $\sim
15$\% of the \CII\ emission, which is similar to our estimate of
10--20\% for the two regions in M\,33.

\subsection{Emission of \CII\ unassociated with CO or \HI}

Use of high spectral resolution observations has led to the detection of
significant \CII\ intensity from molecular gas not detected in CO(2--1).
Such excess \CII\ emission has been observed along many Galactic lines
of sight \citep[][and references therein]{pineda2013}, as well as in
external galaxies including the SMC and LMC \citep[][Requena-Torres et
al in prep.]{madden1997, okada2015}. A possible explanation is that this
excess \CII\ emission is produced in the envelopes of dense molecular
clouds in which the \HI/H$_2$ transitions are largely complete, but the
column densities are low enough such that the \cplus/C$^0$/CO transition
is still far from complete.  \citet{wolfire2010} theoretically estimated
that  the mass fraction of CO-dark molecular  gas increases with
decreasing  $A_{\rm V}$, since relatively more molecular H$_2$ material
lies outside the CO region in this case.  For M\,33 we estimate the
fraction of \CII\ intensity without any corresponding CO(2--1) and \HI\
emission to be between 20--60\% in the central region and below 30\% in
the BCLMP302 region.  This difference in the estimated fraction of
CO-dark molecular gas toward \CII\ emission in two regions of M\,33
could be due to the difference in the FUV flux, extinction, and 
metallicity.  Since within the inner 2\,kpc of M\,33 the metallicity
does not vary significantly \citep{magrini2007}, we compare the FUV
intensities and extinction in the two regions next.

For stars embedded in dusty clouds, the total FUV photon energy
deposited into the cloud is re-emitted in the FIR, hence the FUV
intensity varies linearly as the total FIR continuum intensity
\citep{kaufman1999}. Thus to estimate the FUV intensity we first
estimated the total FIR continuum intensity in the central and BCLMP302
regions from the PACS 70, 100, and 160 micron intensities.  For M\,33,
\citet{xilouris2012} found that a modified blackbody with $\beta = 1.5$
and warm dust temperatures of the order of 55\,K provide the best fit. Using
these parameters Nikola et al. (in prep) estimated that the total FIR
intensities for the two regions in M\,33 studied here have similar
values with peaks of 2--3\,10$^{-15}$\,Watt\,m$^{-2}$\,pixel$^{-1}$ for
a 3\arcsec\ pixel.  The similarity in range of FIR intensities in the two
regions indicate similar values of the far ultraviolet (FUV) intensities
for the two regions.

Based on the 24\,\micron\ MIPS and the H$\alpha$ maps for M\,33, an
extinction map for the entire galaxy was generated by Monica Relano
(priv. comm) via the method presented by \citet{relano2009}. The $A_{\rm
V}$ thus calculated are similar for the central and BCLMP302 regions,
with typical values between 0.6--0.7 mag. At the position of the \CII\
peak in BCLMP302 the value of $A_{\rm V}$ is 0.4\,mag. At the location
of the \CII\ peaks in the central region, $A_{\rm V}$ ranges between
0.5--1.6.  The somewhat higher values of $A_{\rm V}$ seen toward the
center also does not help us explain the larger proportion of CO-dark
molecular H$_2$ at the center. The $A_{\rm V}$ values presented here are
averaged over a 12\arcsec\ beam. Given the inhomogeneous nature of the
ISM, it is likely that the actual value of $A_{\rm V}$ on smaller
spatial scales is significantly larger than the average estimated.

Thus at a resolution of 50\,pc the difference in metallicity, FUV
radiation field, and extinction between the center and the BCLMP302
regions are not significant enough to explain the difference in the
CO-dark component of molecular gas.  This indicates a variation of the
relative amounts of diffuse (CO-dark) and dense molecular gas on spatial
scales smaller than 50\,pc.

\subsection{Interpretation of correlation between \CII, CO, and \HI
\label{sec_conclusion}}

We find that the derived correlation between \CII, CO and \HI\
intensities depends crucially on the availability of spectral and
spatial resolution. Velocity-resolved spectra for BCLMP302 show very
little \CII\ emission not associated with CO and \HI. However, \HI\ and
CO are spatially disjunct and integrated intensities without velocity
information mix \CII\ contributions from \HI\ and CO. Thus, we find a
low correlation between \CII\ and either of them when looking at just
the integrated intensities.  When we are able to filter out CO and \HI\
gas based on the velocity information, we obtain a much better
correlation (Table 1) as we only compare contributions corresponding to
the same velocity ranges.  For the center, we estimate a large amount
of nonassociated \CII\ emission leading to a low correlation when we are
able to determine the associations based on the velocity information, as
reflected by the lower numbers in Table 1. The higher correlation when
considering integrated intensities (Fig.\,\ref{fig_corrplot}) must
indicate a spatial relation of the CO-dark \CII\ emitting molecular gas
to the other tracers. Thus, for the central region, a possible explanation
of the derived correlations could be in terms of production of the
excess \CII\ emission in gas falling onto or outflowing from denser CO
gas. These results show the complementarity of information provided by
observations with or without spectral resolution, which needs to be
factored in when interpreting observed correlations or the lack of it as
seen in galaxies at larger distances.


\section{Summary \label{sec_summary}}

A comparison of velocity-resolved observations of \CII\ with other
dedicated tracers of neutral atomic and molecular gases has been used to
find the relative contribution of the atomic and molecular phases of the
ISM at all positions in two regions within the Triangulum galaxy M\,33.
The difference in the estimated \CII\ emission that is not originating
from CO(2--1) and \HI\ emitting clouds, between the center and the
BCLMP302 regions does not have any obvious explanation in terms of the
variation of extinction, FUV radiation field, and metallicity of the gas
at spatial scales of 50\,pc. This suggests the role of smaller scale
variations of all these parameters toward the observed \CII\ emission.
Based on detailed analysis of one position each in the center and in the
BCLMP302 regions, we estimate that 70\% and 89\% of gas-phase carbon is
in the form of \cplus.  In both these regions at selected locations
coinciding with \HII\ regions, 10--20\% of the \CII\ intensities arise
from the ionized gas. However in the absence of any velocity information
about the ionized gas it is not possible to conclude whether this could
explain at least part of the \CII\ emission not corresponding to the
\HI\ and CO(2--1) emission.  While the velocity information and better
spatial (50\,pc) resolution significantly improves our understanding of
the origin of \CII\ in emission, even from nearby extragalactic sources,
such as  M\,33, where the linear resolution in the present observations
is 50\,pc, the situation is still complicated by the intermixing of
emission from different physical entities within the same beam. 


\begin{acknowledgements}
B.M. acknowledges support received for data reduction from the
Herschel helpdesk, D.  Teyssier, in particular.  V.O.  acknowledges
support by the German Deutsche Forschungsgemeinschaft, DFG project
number Os 177/2-2.
HIFI has been designed and built by a consortium of institutes and
university departments from across Europe, Canada and the US under the
leadership of SRON Netherlands Institute for Space Research, Groningen,
The Netherlands with major contributions from Germany, France and the
US. Consortium members are: Canada: CSA, U.Waterloo; France: CESR, LAB,
LERMA, IRAM; Germany: KOSMA, MPIfR, MPS; Ireland, NUI Maynooth; Italy:
ASI, IFSI-INAF, Arcetri-INAF; Netherlands: SRON, TUD; Poland: CAMK, CBK;
Spain: Observatorio Astron\'omico Nacional (IGN), Centro de
Astrobiolog\'{\i}a (CSIC-INTA); Sweden: Chalmers University of
Technology - MC2, RSS \& GARD, Onsala Space Observatory, Swedish
National Space Board, Stockholm University - Stockholm Observatory;
Switzerland: ETH Z\"urich, FHNW; USA: Caltech, JPL, NHSC.
\end{acknowledgements}

\end{document}